**Long-Range Surface-Assisted Molecule-Molecule Hybridization**

*Marina Castelli, Jack Hellerstedt, Cornelius Krull, Spiro Gicev, Lloyd C.L. Hollenberg, Muhammad Usman[*], Agustin Schiffrin[*]*

Dr. M. Castelli, Dr. J. Hellerstedt, Dr. C. Krull, Dr. A. Schiffrin

School of Physics and Astronomy, Monash University, Clayton, Victoria 3800, Australia

Email: agustin.schiffrin@monash.edu

Dr. M. Castelli, Dr. C. Krull, Dr. A. Schiffrin

ARC Centre of Excellence in Future Low-Energy Electronics Technologies, Monash University, Clayton, Victoria 3800, Australia

S. Gicev, Dr. M. Usman, Prof. L.C.L Hollenberg

Centre for Quantum Computation and Communication Technology, School of Physics, The University of Melbourne, Parkville, 3010, Victoria, Australia

Email: muhammad.usman@unimelb.edu.au

Dr. M. Usman

School of Computing and Information Systems, Melbourne School of Engineering, The University of Melbourne, Parkville, 3010, Victoria, Australia

Abstract: Metalated phthalocyanines (Pc's) are robust and versatile molecular complexes, whose properties can be tuned by changing their functional groups and central metal atom. The electronic structure of magnesium Pc (MgPc) - structurally and electronically similar to chlorophyll – adsorbed on the Ag(100) surface is investigated by low-temperature scanning tunneling microscopy (STM) and spectroscopy (STS), non-contact atomic force microscopy (ncAFM) and density functional theory (DFT). Single, isolated MgPc's exhibit a flat, four-fold rotationally symmetric morphology, with doubly degenerate, partially populated (due to surface-to-molecule electron transfer) lowest unoccupied molecular orbitals (LUMOs). In contrast, MgPc's with neighbouring molecules in proximity undergo a lift of LUMOs degeneracy, with a near-Fermi local density of states with reduced two-fold rotational symmetry, indicative of a long-range attractive intermolecular interaction. The latter is assigned to a surface-mediated two-step electronic hybridization process. First, LUMOs interact with Ag(100) conduction electrons, forming hybrid molecule-surface orbitals with enhanced spatial extension. Then, these delocalized molecule-surface states further hybridize with those of neighbouring molecules. This work highlights how the electronic structure of molecular adsorbates – including orbital degeneracies and symmetries – can be significantly altered via surface-mediated intermolecular hybridization, over extended distances (beyond 3 nm), having important implications for prospects of molecule-based solid-state technologies.

## 1. Introduction

Tetrapyrrole molecules – in particular phthalocyanines (Pc's) and porphyrins – with single coordinated metal atoms at their core allow for a vast range of applications in, e.g., catalysis[1], photovoltaic devices[2], light-emitting devices[3], molecular magnets[4], molecular rotors[5], nanoelectronics[6], gas sensing[7], molecular switches[8]. These functionalities are dictated by the electronic structure at the single molecule level[9]. Systems of practical interest are generally composed of many molecules, and their overall electronic character can be affected by intermolecular interactions (e.g., covalent, non-covalent, electrostatic, magnetic). Moreover, solid-state technologies based on such compounds require interfacing with a solid surface, which can further perturb their electronic properties. It is therefore of fundamental and technological importance to develop an in-depth understanding of the electronic structure of such metal-organic systems, in particular of their interactions with an underlying solid as well as with other adsorbates.

The morphology of neutral, non-interacting metal-porphyrins and metal-Pc's (M-Pc's) is natively planar, with four-fold rotational symmetry, and mirror symmetry planes along the isoindole-isoindole molecular axes and perpendicular to the molecular plane[9b, 9c]. Their electronic structure is characterised by two degenerate lowest unoccupied molecular orbitals (LUMOs) with strong (if not dominant) contributions from the organic Pc ligand[10]. When adsorbed on a solid surface, in some instances, these structural and electronic molecular symmetries can be maintained[4b, 11]. In other cases, these can be broken via anisotropic electronic interactions with the underlying substrate[12] or with other adsorbates[13]. Such a break of symmetry can be accompanied by a lift of LUMOs degeneracy[12d, 13].

Upon adsorption, the doubly degenerate LUMOs can become partially occupied due to surface-to-molecule electron transfer which can lift the degeneracy via Jahn-Teller (JT) distortion, stabilizing the extra negative charge and reducing the rotational symmetry from four- to two-fold[12d, 13a, 14]. On a metal surface, unpaired electrons of partially populated LUMOs, or core metal $d$- orbitals, can result in localized magnetic moments[15] that can either give rise to magnetic anisotropy[4a] and magnetically ordered organic films[16], be quenched via hybridization[17], or be screened by the surface conduction electrons manifesting as a Kondo effect[18].

Scanning tunnelling microscopy (STM) and differential conductance (d$I$/d$V$) scanning tunnelling spectroscopy (STS[19]) allow for addressing and correlating changes in morphology and electronic structure of adsorbed molecules with intramolecular real-space resolution. However, phenomena such as surface-to-molecule charge transfer, Kondo effect, lifts of orbital degeneracies due to JT distortion, and molecular vibrational modes, often give rise to d$I$/d$V$ STS signatures at or near the Fermi level (within an energy window typically on the order of tens of meV[13b, 15, 18a, 18b]). This can make it challenging to disentangle and fully understand these different physical phenomena.

Because of their potential magnetic functionality, adsorbed M-Pc's with *d*-block transition metal atoms at their core (e.g., NiPc, FePc, CuPc, ZnPc, CoPc) have been studied extensively by STM and d$I$/d$V$ STS, in particular on noble metal (100) surfaces, where they preserve their structural and electronic, four-fold rotational symmetry [4b, 12a, 20]. Their electronic structure often reflects surface-to-molecule electron transfer and (elastic and vibrationally assisted) Kondo effects[4b, 12b, 21]. Adsorbed M-Pc's with a simpler electronic structure (e.g., with *s*-block transition metal atoms at their core, lacking *d*-electrons) can make it easier to disentangle these contributions but remain underexplored. An example of such *s*-block M-Pc is magnesium phthalocyanine (MgPc[10]), which is structurally similar to the chlorophyll molecule[22] responsible for photosynthesis in bio-organisms, and can offer new possibilities for developing solid-state light-harvesting and emitting technologies[2c, 23], [24]. Despite its fundamental and technological relevance, a detailed STM/STS characterisation of the atomic-scale morphology and electronic structure of MgPc adsorbed on a noble metal (100) surface has not been performed to our knowledge.

Here we show by means of low-temperature STM, d$I$/d$V$ STS, non-contact atomic force microscopy (ncAFM) and density functional theory (DFT), that single MgPc molecules on Ag(100) are four-fold rotationally symmetric with doubly degenerate and partially populated LUMOs, similar to other M-Pcs[4b, 12a, 13b, 18a, 21]. However, when another MgPc is in proximity (for intermolecular distances up to ~3 nm), the LUMOs degeneracy can be lifted and the near-Fermi LDOS can exhibit a reduced two-fold rotational symmetry. We explain our observations by a two-step electronic hybridization process resulting in an effective long-range attractive molecule-molecule interaction, where: (i) the degenerate LUMOs of an MgPc molecule interact with Ag(100) conduction electron states, forming hybrid molecule-surface orbitals with enhanced spatial extension; (ii) these delocalized hybrid molecule-surface orbitals associated with a pair of nearest-neighbouring molecules can hybridize in turn. This work reports a first experimental evidence and detailed theoretical understanding of long-range substrate-mediated molecule-molecule electronic hybridisation, resulting in an effective attractive intermolecular interaction which survives beyond distances of ~3 nm. Such interaction could have important implications in the design of prospective molecule-based solid-state technologies.

2. Results

2.1 Structural Characterization of MgPc on Ag(100): STM, ncAFM and DFT

We deposited MgPc molecules onto a clean Ag(100) surface via thermal evaporation (see Methods). We subsequently characterised the MgPc/Ag(100) system via low-temperature (4.6 K) STM. **Figure 1a** shows an STM topographic image of molecules appearing as cross-like features, characteristic of the phthalocyanine (Pc) ligand[25]. The molecules adsorb with their isoindole-isoindole axes (green and yellow dashed lines in Figure 1a) following a ±29 ± 1° angle with respect to the [011] crystalline directions of Ag(100), consistent with previous studies[4b, 12a]. We identified two types of MgPc's on the Ag(100) surface. The first type labelled "single" has the four-fold rotational symmetry and mirror symmetry planes along the isoindole-isoindole axes inherited from the gas phase, as shown in the apparent height profiles (solid yellow and green curves) in Figure 1a. The second type consists of MgPc's whose low-bias STM topography, in contrast with the "single" molecules, reveals a reduced two-fold rotational symmetry, with two mirror-symmetric isoindole groups having an apparent height larger than those along the other orthogonal isoindole-isoindole axis [see purple height profile difference (purple area) in Figure 1a]. Notably, we observed this reduction of symmetry for MgPc's that are in proximity of another MgPc molecule, with their lowest-apparent-height isoindole-isoindole axis collinear to the axis defined by the two Mg-Mg centers. We named such pairs of nearest-neighbour MgPc's with collinear isoindole-isoindole axis "++" dimers (orange frame in Figure 1a).

In order to determine whether the reduction of rotational symmetry observed in the STM topographies is due to a change in intramolecular morphology, we performed frequency-shift ($\Delta f$) ncAFM measurements with a carbon-monoxide (CO)-functionalized tip (Methods), on a single MgPc (Figure 1b) and a "++" dimer (Figure 1c). This technique enables real-space intramolecular structural characterisation with chemical bond resolution[26]. Both the single molecule and "++" dimer cases show very similar four-fold rotationally symmetric morphologies, with the four distinguishable isoindole units identical to their Lewis structure. The Mg centers of both configurations show a "hash"-shaped appearance, where the four nitrogen (N) atoms of the pyrrole rings seem interconnected. The latter could be explained by the flexibility and dipole moment of the CO tip, known to introduce artefacts[27]. Similar imaging was reported for FePc[28], attributed to a lowered density of electrons at the metal centre. Importantly, constant-height $\Delta f$ profiles along the isoindole-isoindole axes (solid green and yellow curves in Figure 1b and c) corroborate that the intramolecular morphology of MgPc remains the same (within our experimental resolution) for both single and "++" dimer configurations.

We performed DFT calculations to determine the theoretical lowest-energy (relaxed) morphology of the single isolated MgPc and of the "++" dimer on the Ag(100) surface (Figure 1d, e; see Methods for computational details). These calculations show that, in both cases, the intramolecular conformation is quasi-identical, with the central Mg atom sitting on top of a Ag(100) hollow site, and with molecular height differences of, at most, ~3 pm (smaller than our experimental error). This agrees with our STM and ncAFM data [Figure 1b, c, and Figure S1, S2 in Supporting Information (SI)], and with previously reported studies of other M-Pc's on (100) noble metal surfaces[4b, 12a, 20]. We conclude that the symmetry reduction observed in Figure 1a is not due to a different adsorption site of the molecule, nor to an intramolecular structural deformation.

To quantify the break of molecular symmetry observed via STM (Figure 1), we defined a topographic asymmetry parameter, $\alpha_{STM} = \frac{|(z_1+z_3)-(z_2+z_4)|}{z_1+z_3+z_2+z_4}$, where $z_j$ ($j = 1,…, 4$) is the maximum STM apparent height of each of the four isoindole units of a molecule (such that $z_1 > z_3 > z_2 > z_4$). Figure 1f shows $\alpha_{STM}$ for a single isolated MgPc (blue) and for MgPc of a "++" dimer (Mg-Mg distance $d_{c-c} = 2.1$ nm; orange) as a function of applied bias voltage $V_b$. For the single MgPc, $\alpha_{STM}$ is close to zero and independent of $V_b$. The "++" dimer is qualitatively different, with $\alpha_{STM}$ peaking at $V_b = 0.02$ V and attenuating symmetrically (see SI Figure S3). Combined with the ncAFM imaging and DFT calculations of the intramolecular morphology, we conclude from the bias voltage dependence of $\alpha_{STM}$ that the observed break of STM topography symmetry is exclusively the result of changes in the molecular near-Fermi electronic structure induced by interactions with a neighbouring molecule.

**2.2 Electronic Structure of MgPc on Ag(100): d$I$/d$V$ STS and DFT**

To elucidate the electronic structure of both single MgPc and MgPc in "++" dimers, we performed differential conductance (d$I$/d$V$) STS measurements (Methods). **Figure 2a** contains d$I$/d$V$ point spectra taken at the periphery of the Pc ligand (blue) and on the Mg center (red), for both single MgPc (triangles) and "++" dimer (squares). In both single MgPc and dimer cases, we identified a prominent feature at $V_b$ = -1.45 V. The spatial distribution (see d$I$/d$V$ maps in Figure 2b, d) of this resonance is quasi-identical for both cases. It is four-fold rotationally symmetric and exclusively located on the peripheral Pc moiety, with two nodal planes normal to the Ag surface along each isoindole-isoindole axes (grey arrows), and with no contribution from the Mg center. Consistently with previous studies[4b] of other M-Pc's on

Ag(100), we attribute this resonance to the gas-phase highest occupied molecular orbital (HOMO). This HOMO spatial distribution is also slightly chiral (i.e., no mirror symmetry), which we attribute to inhomogeneous interactions with the substrate[4b, 12a]. Figure 2e, g show simulated differential conductance maps derived from DFT-calculated orbitals (Methods), for a single MgPc and an MgPc "++" dimer on Ag(100), for $V_b \cong -1.4$ V, in agreement with our experimental d$I$/d$V$ maps in Figure 2b, d. Given the four-fold rotational symmetry, strong spatial resemblance between single molecule and dimer cases, and the corresponding energy significantly far away from that at which $\alpha_{STM}$ is maximum (Figure 1f), we conclude that the HOMO plays no role in the break of molecular topographic symmetry observed in Figure 1.

The d$I$/d$V$ spectra acquired at the center of the molecule (solid red curves in Figure 2a) are dominated by a resonance at 2.8 V, for both single MgPc and "++" dimer. The d$I$/d$V$ map in Figure 2c shows the spatial distribution of this resonance, mainly localized on the Mg atom and the inner pyrrole rings, consistent with our DFT calculation (Figure 2f). We attribute this resonance[10] to higher energy unoccupied states of the gas-phase molecule (additional details in SI section S13). Given this central spatial distribution and corresponding energy of 2.8 V, far above the Fermi level, we conclude that the electronic state associated with this feature is not involved in the near-Fermi STM topographic asymmetry in Figure 1, which is mainly related to the outer isoindole groups.

At small absolute values of bias voltage, d$I$/d$V$ spectra at the Mg center, for both the single MgPc and "++" dimer, show step-like features (red arrows in Figure 2a) symmetric in energy with respect to the Fermi level ($V_b = 0$), and a sharp resonance at ~ -0.1 V on the Pc ligand. In **Figure 3** we focus on these near-Fermi spectroscopic features. Note that we did not observe any Kondo-effect-related zero bias peak or dip. We fit (Figure 3a) the central Mg-related spectra with a sum of an attenuated reference spectrum taken on bare Ag(100) (to account for tip features), and three pairs of Fermi-Dirac distributions symmetric with respect to $V_b = 0$, with onsets at $\pm 0.005 \pm 0.001$, $\pm 0.089 \pm 0.001$ and $\pm 0.2 \pm 0.001$ V (see SI section S6 for details on fitting). Given their step-like near-Fermi character and their symmetry with respect to the Fermi level, we attribute these features to inelastic contributions to the differential conductance given by molecular vibrational modes associated with the Mg-N bonds and stretching of the isoindole units. This is consistent with previous studies of MgPc and other M-Pc's[18a, 29]. The d$I$/d$V$ maps at $V_b = \pm 0.2$ V in Figure 3b-e indicate that the spatial distributions associated with these step-like features are four-fold rotationally symmetric and quasi-identical for both single and "++" dimer cases.

The blue curve in Figure 3f corresponds to the d$I$/d$V$ spectrum at the peripheral Pc ligand of a single MgPc (averaged over areas indicated by dashed blue circles in Figure 3g). Similar to the Mg center, we fit this d$I$/d$V$ spectrum with a pair of energy-symmetric Fermi-Dirac distributions (onsets at $\pm 0.202 \pm 0.001$ V; related to the most prevalent vibrational mode) and a Gaussian peak (centered at $-0.097 \pm 0.001$ V). The d$I$/d$V$ map associated with this Gaussian peak (Figure 3g) is four-fold rotationally symmetric. We attribute this Gaussian peak to the doubly degenerate gas-phase LUMOs, where each LUMO extends along two opposite isoindole groups, with nodal planes (indicated by grey arrows in Figure 3g) orthogonal to the molecular plane and along the isoindole-isoindole axes [25c] (see SI Figure S17). This is consistent with the simulated d$I$/d$V$ maps derived from DFT calculations (Methods), at an energy related to the MgPc/Ag(100) LUMOs (Figure 3k). The Gaussian peak associated with the LUMOs crosses the Fermi level, indicative of partial filling of the LUMOs due to Ag(100)-to-MgPc electron transfer. Bader charge analysis[30] based on our DFT calculations estimates that ~1.3 electrons are transferred to MgPc upon adsorption, in agreement with previous work[4b]. The four-fold rotational symmetry of this d*I*/d*V* map indicates that the transferred negative charge populates each of the two LUMOs with equal probability, and that adsorption on Ag(100) does not lift the LUMOs degeneracy.

In the "++" dimer configuration, the Pc ligand d*I*/d*V* spectra (orange and green curves in Figure 3f) are qualitatively similar to that for single MgPc. However, they present subtle differences depending on the exact acquisition location (orange and green dashed circles in Figure 3i, j). Whilst we were able to fit well the orange spectrum in Figure 3f with a pair of Fermi-Dirac distributions (onsets at $\pm 0.198 \pm 0.001$ V; vibrational mode) and a Gaussian peak ($-0.098 \pm 0.001$ V), the green spectrum required the addition of an extra Gaussian peak at $0.157 \pm 0.002$ V (see SI Section S6). Figure 3i, j show d*I*/d*V* maps associated with these Gaussian peaks, at -0.08 and 0.15 V, respectively. The map at 0.15 V was subtracted by the d*I*/d*V* map at 0.08 V (i.e., $\Delta(dI/dV)_{(0.15\,V;\,0.08\,V)} = (dI/dV)_{0.15\,V} - (dI/dV)_{0.08\,V}$), in order to attenuate inelastic contributions from molecular vibrational modes (which dominate the spectra near Fermi; Figure 3c, e) and enhance peripheral Pc ligand features related to the Gaussian peak. These d*I*/d*V* maps exhibit two-fold rotational symmetry, with only one nodal plane parallel (perpendicular) to the Mg-Mg dimer axis at -0.08 V (0.15 V); see Figure 3i (Figure 3j, respectively). The d*I*/d*V* map at -0.08 V of a molecule in the "++" dimer, rotated by 90°, is qualitatively identical to that at 0.15 V for the same molecule. This is in stark contrast to the four-fold rotationally symmetric single MgPc case (Figure 3g). We attribute the two Gaussian peaks in the green d*I*/d*V* spectrum in Figure 3f, and their associated reduced spatial symmetry (from four-fold rotational to two-fold rotational; Figure 3i, j), to the two LUMOs, with their degeneracy lifted by interactions with the neighbouring molecule in the dimer. One LUMO is partially occupied (-0.08 V); the other is empty (0.15 V). This reduction of orbital rotational symmetry and lift of LUMOs degeneracy is the cause of the break of symmetry observed in the STM topography in Figure 1. This is consistent with DFT-simulated d*I*/d*V* maps of the "++" dimer (Figure 3m, n), which reveal near-

Fermi occupied and empty electronic states, associated with the LUMOs, and with two-fold rotationally symmetric spatial distributions orthogonal to each other.

To address the influence of the noble metal substrate, we used DFT to calculate differential conductance maps of a neutral MgPc "++" dimer in the gas phase, at energies related to the (here unoccupied) LUMOs (see SI Figure S19). We observed a similar lift of degeneracy and reduction of rotational symmetry of the LUMOs, but only for small ($d_{c-c} < 2$ nm) Mg-Mg intermolecular distances. We therefore infer that the extended spatial range of the intermolecular interaction on Ag(100) is surface mediated.

Similar phenomena have been observed previously for other M-Pc's with partially filled LUMOs (i.e., negatively charged) on metals and atomically thin insulators, where degeneracy lifting and rotational symmetry reduction were induced by Jahn-Teller distortions mediated by inhomogeneous or anisotropic interactions with the substrate [12d, 25c] or neighbouring adatoms[13a]. In the latter case, the molecule-adatom interaction was repulsive, resulting in the lowest-energy LUMO nodal plane being perpendicular to the axis defined by the M-Pc center and the adatom. In our "++" dimer case, we observe the opposite: the lowest-energy LUMO nodal plane is parallel to the Mg-Mg dimer axis (Figure 3i), whereas the nodal plane of the high energy LUMO (Figure 3j) is perpendicular to this axis. This indicates that the interaction between the MgPc molecules in the "++" dimer is effectively attractive (see SI Figure S16), despite their identical (negative) charge.

The d$I$/d$V$ map at 0.15 V for a single isolated MgPc (Figure 3h; subtracted by the 0.08 V d$I$/d$V$ map) is four-fold rotationally symmetric, with two orthogonal nodal planes, similar to the d$I$/d$V$ map at -0.08 V related to the degenerate LUMOs (Figure 3g). We understand the similarity between the -0.08 and 0.15 V maps as a consequence of the partial filling of the doubly degenerate LUMOs, and of the Hubbard energy $U$ necessary to overcome the Coulomb repulsion when injecting a tunneling electron into these[18a]

(i.e., for $V_b > 0$). The 0.15 V map can be interpreted as a residue (tail) of a possible LUMO+$U$ spectroscopic feature at higher energy (see SI Section S9).

To quantify the reduction of LUMOs rotational symmetry observed for MgPc in a "++" dimer, we defined from our d$I$/d$V$ maps at $V_b$ = -0.08 V (e.g., Figure 3g, i) an experimental spectroscopic asymmetry parameter $\alpha_{dI/dV}$:

$$\alpha_{dI/dV} = \frac{\iint |\mathcal{R}_{90°}[(dI/dV)_{bin}(x, y, V_b = -0.08 \text{ V})] - (dI/dV)_{bin}(x, y, V_b = -0.08 \text{ V})| \, dx \, dy}{\iint |\mathcal{R}_{90°}[(dI/dV)_{bin}(x, y, V_b = -0.08 \text{ V})] + (dI/dV)_{bin}(x, y, V_b = -0.08 \text{ V})| \, dx \, dy}$$

where $(dI/dV)_{bin}(x, y, V_b = -0.08 \text{ V})$ is the binarized d$I$/d$V$ map as a function of tip position $(x, y)$, and $\mathcal{R}_{90°}[(dI/dV)_{bin}(x, y, V_b = -0.08 \text{ V})]$ its 90° clockwise rotation around an axis perpendicular to the surface going through the Mg center (see SI section S11 for details). We calculated $\alpha_{dI/dV}$ for MgPc in "++" dimers with different Mg-Mg intermolecular distances $d_{c-c}$ (**Figure 4a**), yielding $\alpha_{dI/dV} = 1$ for the dimer with the smallest $d_{c-c}$ considered (~2 nm, i.e., most marked two-fold rotational symmetry), decaying monotonically until $\alpha_{dI/dV} \to 0$ for $d_{c-c} > 3$ nm (i.e., four-fold rotational symmetry is recovered, as for single isolated MgPc).

Similarly, we calculated theoretical asymmetry parameters $\alpha_{DFT}^{(Ag)}(d_{c-c})$ and $\alpha_{DFT}^{(GP)}(d_{c-c})$ (red triangles and blue stars in Figure 4a) from DFT-simulated differential conductance maps for "++" dimers on Ag(100) and in the gas phase (GP), respectively, at a similar bias voltage related to the low energy LUMO (Figure 3m; see SI Figure S19 for corresponding DFT-calculated d$I$/d$V$ maps). The dependence of $\alpha_{dI/dV}$, $\alpha_{DFT}^{(Ag)}$ and $\alpha_{DFT}^{(GP)}$ on $d_{c-c}$ can be fit with a decaying exponential function $\propto e^{-d_{c-c}/\lambda}$ (dashed fit curves in Figure 4a), with $\lambda_{dI/dV} = 0.5 \pm 0.2$ nm and $\lambda_{DFT}^{(Ag)} = 0.6 \pm 0.1$ nm, showing good agreement between experiment and theory. For the gas phase "++" dimer, $\lambda_{\text{DFT}}^{(\text{GP})} = 0.09 \pm 0.02$ nm. That is, the range of the effectively attractive intermolecular interaction is significantly larger (> 5 times) on the noble metal surface than in the gas phase. Whilst $\alpha_{dI/dV}$, $\alpha_{\text{DFT}}^{(\text{Ag})} > 0$ for $d_{c-c} > 3$ nm, $\alpha_{\text{DFT}}^{(\text{GP})} \cong 0$ for $d_{c-c} \cong 2$ nm (see details in SI Figure S19).

3. Discussion

We considered several phenomena to explain the effective intermolecular attractive interaction that reduces the LUMO rotational symmetry observed both experimentally and in our DFT calculations. We exclude dipole-dipole or substrate-mediated (Friedel oscillations, spin-spin) interactions as lacking consistency with our observations (see SI section S12 for details). Our DFT calculations of single MgPc on Ag(100) show significant spectral energy broadening and spatial delocalization of the LUMOs (when compared to those of the gas phase), due to hybridization with the substrate conduction electron states (SI Figure S17). Motivated by our evidence that the effective intermolecular interaction between MgPc's in a "++" dimer is attractive, we considered a simple model where we assume that: (i) the MgPc negative charge resulting from the LUMOs partial filling via surface-to-molecule electron transfer is screened by the Ag(100) conduction electrons, effectively cancelling intermolecular Coulomb repulsion (note that attractive interactions between molecules with charge of same sign on a metal have been observed previously[31]); and (ii) two spatially delocalized orbitals (each resulting from the interaction between a LUMO and Ag(100) conduction electrons, and each associated with a different MgPc in a "++" dimer) can in turn hybridize, splitting the initially two-fold degenerate LUMOs into non-degenerate ones. In this model, for a "++" dimer with a specific $d_{c-c}$, we considered four (two per MgPc) degenerate delocalized LUMO/Ag(100)-conduction-electron orbitals, that we first assumed non-interacting (i.e., Figure 4d). We then considered an effective one-electron Hamiltonian including two attractive central potentials ($\propto -1/r$) centered at each of the two Mg atoms (accounting for attractive interactions between an electron and the molecules' nuclei; see SI section S14), and constructed new hybrid orbitals for the interacting "++" dimer via linear combination of the four, initially degenerate molecular orbitals (LCMO). We used the two lowest energy LCMO-generated hybrid orbitals (labelled $|\varphi_1\rangle$ and $|\varphi_2\rangle$) to simulate the corresponding differential conductance maps (Figure 4b, c; see Methods and SI Figure S19). Up to $d_{c-c} \cong 3$ nm, these maps exhibit a break of four-fold rotational symmetry, with a nodal plane parallel (perpendicular) to the Mg-Mg dimer axis for $|\varphi_1\rangle$ ($|\varphi_2\rangle$, respectively), consistent with our experimental (Figure 3i, j) and DFT-calculated (Figure 3m, n) d$I$/d$V$ maps.

These LCMO-derived differential conductance maps allowed us to calculate a spectroscopic asymmetry parameter $\alpha_{\text{LCMO}}^{(\text{Ag})}$ as a function of $d_{c-c}$ (red hashes, Figure 4a). Exponential fitting of $\alpha_{\text{LCMO}}^{(\text{Ag})}(d_{c-c})$ yielded $\lambda_{\text{LCMO}}^{(\text{Ag})} = 0.5 \pm 0.2$ nm, in close agreement with the experimental $\alpha_{dI/dV}(d_{c-c})$ and DFT-derived $\alpha_{\text{DFT}}^{(\text{Ag})}(d_{c-c})$ (see $\lambda_{dI/dV}$ and $\lambda_{\text{DFT}}^{(\text{Ag})}$ above). Using DFT-calculated LUMOs of gas phase MgPc as inputs to our LCMO model, we also performed the same procedure to calculate $\alpha_{\text{LCMO}}^{(\text{GP})}(d_{c-c})$ (blue hashes in Figure 4a) that closely matches $\alpha_{\text{DFT}}^{(\text{GP})}$.

The quantitative agreement – for both d$I$/d$V$ maps (Figure 4b, c) and $d_{c\text{-}c}$-dependence of the spectroscopic asymmetry parameter $\alpha$ (Figure 4a) – between experiments, DFT and LCMO model validates the latter. This provides compelling evidence that the physical mechanism behind the observed LUMOs' symmetry reduction and lift of degeneracy consists of an intermolecular hybridization-like process, with an increased spatial range (in comparison to the gas phase) given by the molecule-surface interaction and resulting LUMOs' delocalization.

The intermolecular hybridization should lower the energy of the molecular electronic system and further stabilize (in addition to screening by the metal substrate) the negative charge partially occupying the LUMOs. That is, the eigenenergy of the lowest-energy hybrid bonding orbital $|\varphi_1\rangle$ should be smaller than that of the degenerate LUMOs of non-interacting molecules. However, experimentally we observe no appreciable difference between the energy of the single isolated MgPc LUMOs and that of the "++" dimer $|\varphi_1\rangle$ (see Figure 3f). We attribute this to pinning of the LUMO just below the Fermi level given by the dipole formed at the interface between molecule and Ag(100) surface[32].

Substrate-mediated intermolecular hybridization has been reported previously for a binary closed-packed monolayer of ZnPc's on Ag(111)[33] where Zn-Zn distances are less than 2 nm, and hybrid ZnPc/Ag(111) orbitals were observed. In our case, substrate-mediated intermolecular hybridization clearly persists beyond Mg-Mg distances of ~3 nm, as shown by $\alpha_{dI/dV}$ (Figure 4a). Other STM and d$I$/d$V$ STS studies on 3$d$ TM-Pc's on Ag(100) have not reported any reduction of spatial symmetry of frontier orbitals due to intermolecular interactions[4b]. The near-Fermi electronic structure of these systems is often dominated by features related to the (fully or partially occupied) transition metal $d$-orbitals[17a] (e.g., zero-bias Kondo effect resonance), which can potentially hide or hinder the effect that we observed here. We claim that the MgPc/Ag(100) system, with its simpler electronic structure (only $s$ and $p$ molecular electrons; no near-Fermi d$I$/d$V$ substrate features, e.g., Shockley surface state), and especially with its degenerate, partially populated LUMOs localized at the Pc periphery, allows us to isolate the long-range substrate-mediated intermolecular interaction (and resulting LUMOs lift of degeneracy and symmetry reduction) from other effects. Notably, $\alpha_{dI/dV}$ represents a highly sensitive real-space observable that (compared to, e.g., an energy-dependent LDOS measurement) lets us quantify the spatial dependence of this phenomenon.

## 4. Conclusion

We have studied the near-Fermi electronic structure of MgPc molecules on Ag(100). Our STM, d$I$/d$V$ STS and ncAFM measurements, supported by DFT calculations, show that the LUMOs of single isolated molecules are partially populated and doubly degenerate, resulting in a four-fold rotationally symmetric near-Fermi occupied LDOS. An effective attractive intermolecular interaction between adsorbed MgPc's separated by up to ~3 nm can lift the LUMOs degeneracy and break the rotational spatial symmetry of the near-Fermi molecular states. We explain this interaction within a first-order simplified model (LCMO) that accounts for an effective long-range hybridization between partially populated LUMOs, delocalized due to the adsorption on the noble metal surface. Our study highlights how interactions between molecules on metal surfaces can extend over several nm's due to delocalization of molecular electronic states, significantly altering energy degeneracies and spatial symmetries of the latter. This can have severe effects on (and provide opportunities for control of) electronic properties of molecule-metal interfaces in solid-state systems where functionality depends on such symmetries, such as dipole matrix elements in optoelectronics or non-equilibrium Green's functions in molecular electronics.[34]

## 5. Experimental and Theoretical Methods

*Sample preparation:*

MgPc molecules (Sigma Aldrich) were deposited in ultrahigh vacuum (UHV) from the gas-phase (sublimation temperature: 340°C) onto a clean Ag(100) surface (Mateck GmbH) maintained at room temperature. The Ag(100) was prepared by repeated cycles of Ar$^+$ sputtering and annealing (450°C). Molecules were deposited at sub-monolayer coverages (~10% of 1 monolayer). The base pressure was below 3 x 10$^{-9}$ mbar during depositions.

*STM & dI/dV STS measurements:*

All STM and d$I$/d$V$ STS measurements were performed at 4.6 K in UHV (< 1 x 10$^{-10}$ mbar) with an Ag-terminated Pt/Ir tip. Topographic STM images were acquired in constant-current mode. Differential conductance d$I$/d$V$ spectra (Figure 2a, 3a, 3f) were obtained by averaging multiple $I(V)$ curves (at least 10) and numerical derivation. The d$I$/d$V$ map in Figure 2c was acquired with a lock-in amplifier, in constant-current mode ($V_b$ = 2.8 V, $I_t$ = 250 pA), with a bias voltage modulation amplitude of 20 mV and a frequency of 665 Hz. All other d$I$/d$V$ maps were acquired using a multi-pass (MP) approach (see SI Section S7). This technique consists of: (i) acquiring a constant-current STM topographic profile along a scanning line (we used a setpoint $V_b$ = -2.5 V, $I_t$ = 100 pA, at a scanning speed of 3 nm s$^{-1}$), and (ii) recording d$I$/d$V$ with a lock-in amplifier (see parameters above) while scanning (speed: 1.5 nm s$^{-1}$) the same line and following the same constant-current STM topographic profile as in (i), with the tip approached of an additional 150 pm towards the sample; (iii) finally, repeating this procedure sequentially for each scanned line of the map. This approach has the benefit of minimizing variations of d$I$/d$V$ due to variations in STM apparent topography.

*CO functionalisation:*

All of our ncAFM imaging data (and STM in SI Figure S1) were acquired with a tip (Pt/Ir) functionalized with a carbon monoxide (CO) molecule at its apex. Such tip functionalization was achieved by dosing CO gas into the UHV chamber (5 x 10$^{-8}$ mbar for ~10 s) with the Ag(100) sample held at 8 K, placing the tip above a CO molecule on bare Ag(100) with an STM bias voltage of 3 mV, and approaching the tip towards the surface (feedback off) until the tunneling current reaches ~5 nA and then decreases suddenly due to a CO molecule being picked-up.

*NcAFM measurements:*

Non-contact AFM measurements were performed at 4.6 K in UHV (< 1 x 10$^{-10}$ mbar) using a qPlus tuning fork, in frequency modulation mode (resonance frequency $f \cong 29$ kHz; spring constant $k \cong 1800$ N m$^{-1}$), with a CO-functionalized Pt/Ir tip. Frequency-shift ncAFM maps (Figure 1b, c) were acquired at constant height, with a 60 pm amplitude oscillation, and with the tip approached 40 pm towards the surface with respect to the STM setpoint $V_b = 0.02$ V, $I_t = 5$ pA on bare Ag. No bias voltage was applied during ncAFM map acquisition. NcAFM maps were smoothed by convolution with a Gaussian function, followed by a Laplace edge detection and a minimum filter.[26a]

*DFT calculations:*

We performed DFT calculations for gas-phase MgPc and MgPc on Ag(100) using the SIESTA[35] simulation tool with periodic boundary conditions, Troullier-Martins norm-conserving pseudopotentials (with relativistic corrections[36] for Ag) and the RPBE exchange-correlation[37] functional with van der Waals corrections (Grimme's method)[38]. Kohn-Sham orbitals were represented by a DZP basis set with an energy shift of 0.01 and 0.02 Ry for the gas phase and Ag(100), respectively. The total electron charge density was represented on a real space grid corresponding to a mesh cut-off of 300 Ry. The Ag(100) surface was modelled with a 5-layer slab of Ag atoms and a vacuum gap of 7 layers. The behaviour of a single adsorbed MgPc molecule was modelled on a $8 \times 8$ lateral supercell, while lateral supercells of $11 \times 9$ and $13 \times 9$ were used for MgPc "++" dimer systems. Reciprocal space $k$-points were sampled on a $1 \times 1 \times 1$ Monkhorst-Pack grid[39] for all slab systems. A lattice constant of 4.16 Å was used for bulk

Ag. Structural optimization was performed for a single MgPc (Figure 2) and for a "++" dimer with $d_{c-c}$ ≅ 2.0 nm (Figure 3), by relaxing the atoms of molecules and of the Ag(100) upper 3 layers, such that the Cartesian components of forces on relaxing atoms were reduced below 0.02 eV/Å. Gas-phase "++" dimers (used to calculate $\alpha_{DFT}^{(GP)}$ in Figure 4a) were structurally relaxed with the positions of center Mg atoms fixed. Subsequent calculations for "++" dimers adsorbed on Ag(100) with other values of $d_{c-c}$ (used to calculate $\alpha_{DFT}^{(Ag)}$ in Figure 4a) were performed by assuming a flat molecular structure, with $d_{c-c}$ varied by moving MgPc molecules to different hollow sites of the substrate (maintaining the same adsorption angle). These systems were constructed using the atomic positions of gas-phase MgPc and bare Ag(100), with adsorption heights and substrate interlayer spacings set to their respective averages found from the fully relaxed adsorbed "++" dimer with $d_{c-c}$ ≅ 2.0 nm, without further structural relaxation. This is justified by the negligible distortions of the computed relaxed structure of single MgPc and "++" dimer, in agreement with experimental ncAFM imaging (Figure 1b-e).

We used the DFT-calculated one-electron Kohn-Sham orbitals $\psi_v$ (with eigenenergy $E_v$) to simulate theoretical differential conductance (d$I$/d$V$) STS maps (Figure 2e-g, 3k-n; S15c, S19 in SI) according to the Tersoff-Hamann approximation, where first-order perturbation theory is applied with the assumption of spherically symmetric tip states[40]:

$$\frac{\partial I}{\partial V}(x, y, z_0, V) \propto \sum_v |\psi_v(\vec{r}_0)|^2 \cdot \delta(E_v - E_f + eV)$$

where $\psi_v(\vec{r}_0)$ is a single-electron Kohn-Sham wavefunction with eigenenergy $E_v$, $\vec{r}_0 = (x, y, z_0)$ is the position of the tip ($z_0$: tip-sample distance, where $z = 0$ corresponds to the molecular plane; $xy$ plane is parallel to surface and molecular plane), $V$ is the bias voltage of interest and $E_f$ the Fermi energy.

We first numerically evaluated the single-electron Kohn-Sham orbitals $\psi_v(x, y, z_0)$ at a reference $xy$ plane sufficiently close to atoms such that wavefunctions are defined within numerical error, yet sufficiently far such that the potential does not diverge nor varies steeply. We chose a distance of $z_0 = 1.7$ Å above the molecular plane. We then Fourier expanded $\psi_v(x, y, z_0)$ within this reference plane (consistent with the DFT calculations periodic boundary conditions):

$$\psi_{n,\vec{k}}(x, y, z_0) = \sum_{\vec{G}} a_{\vec{k},\vec{G}}(z_0) \exp(i[(k_x + G_x)x + (k_y + G_y)y])$$

and substituted this Fourier expansion into the vacuum Schrödinger equation[41]:

$$-\frac{\hbar^2}{2m} \nabla^2 \psi_{n,\vec{k}}(x, y, z) = (E_{n,\vec{k}} - E_{\text{vac}}) \psi_{n,\vec{k}}(x, y, z)$$

where $E_{\text{vac}}$ is the vacuum energy, yielding

$$a_{\vec{k},\vec{G}}(z) = a_{\vec{k},\vec{G}}(z_0) \exp\left(-\lambda_{\vec{k},\vec{G}}(z - z_0)\right)$$

with

$$\lambda_{\vec{k},\vec{G}} = \sqrt{\left(\vec{k} + \vec{G}\right)^2 - \frac{2m}{\hbar^2}(E_{n,\vec{k}} - E_{\text{vac}})}$$

where $\vec{k} = (k_x, k_y)$ and $\vec{G} = (G_x, G_y)$. This allows us to evaluate $\psi_{n,\vec{k}}(x, y, z_0)$ at any height $z_0$ with respect to the reference plane. The theoretically calculated d$I$/d$V$ maps shown throughout main text and SI (as well as the derived spectroscopic asymmetry parameters $\alpha_{\text{DFT}}^{(\text{Ag})}, \alpha_{\text{DFT}}^{(\text{GP})}, \alpha_{\text{LCMO}}^{(\text{Ag})}, \alpha_{\text{LCMO}}^{(\text{GP})}$) correspond to a distance of 4.3 Å above the molecular plane (i.e., 2.6 Å above the aforementioned reference plane).

**Supporting Information**

Supporting Information is available from the Wiley Online Library or from the author.

**Acknowledgements**

A.S. acknowledges support from the Australian Research Council (ARC) Future Fellowship scheme (FT150100426). M.C. acknowledges support from the Monash Centre of Atomically Thin Materials (MCATM), and the ARC Centre of Excellence in Future Low-Energy Electronics Technologies (FLEET). The computational resources for this work were provided by the National Computing Infrastructure (NCI) and Pawsey Supercomputing Center through National Computational Merit Allocation Scheme (NCMAS). S.G., L.H. and M.U. acknowledge useful discussions with Charles Hill (The University of Melbourne).

*Author Contributions:*

M.C., J.H., C.K. and A.S. designed and performed all experiments, and analysed and interpreted all data. M.U. supervised the theoretical calculations with input from L.C.L.H. S.G. performed the DFT calculations and theoretical simulations of d$I$/d$V$ STS maps. M.C. performed the LCMO calculations with input from S.G. M.C., J.H., S.G., M.U. and A.S. wrote the manuscript. All authors discussed the results and contributed to the final manuscript.

**Conflict of Interest**

The authors have no financial or commercial conflicts of interest to declare.

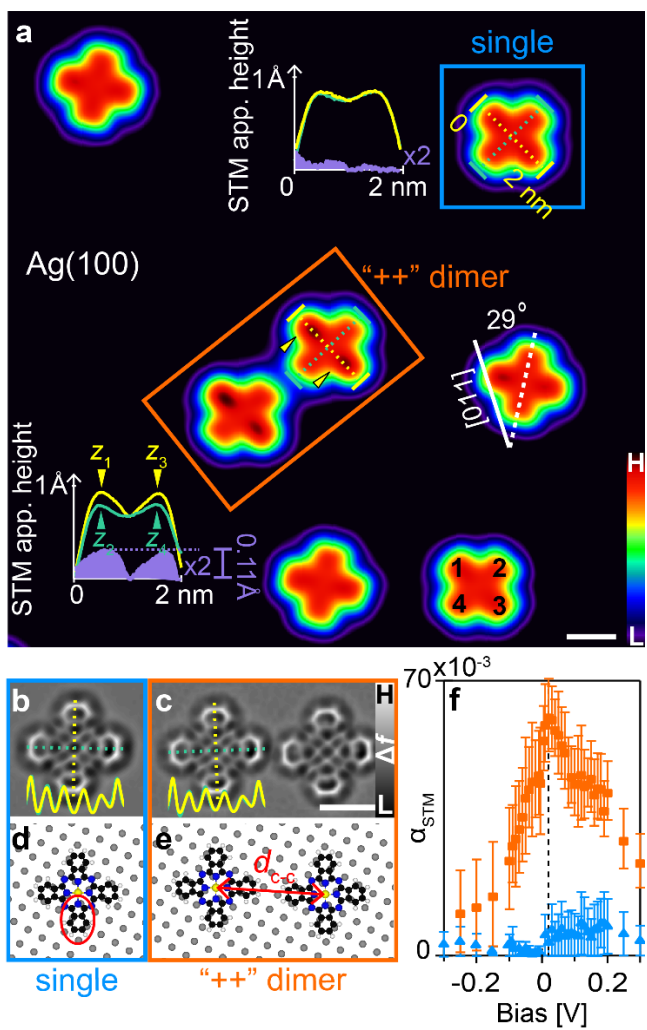

**Figure 1. Symmetry reduction of STM apparent topography induced by nearest-neighbour MgPc on Ag(100).** (a) Constant-current STM image ($V_b$ = 0.02 V, $I_t$ = 5 pA) of MgPc molecules on Ag(100). Single isolated molecules (blue frame) show four-fold rotational symmetry. Molecules with a neighbouring MgPc, with collinear isoindole-isoindole axes ("++" dimer; orange frame; Mg-Mg distance: $d_{c-c}$ = 2.29 nm) appear two-fold symmetric. Insets: STM apparent height profiles (solid yellow and green curves), along the isoindole-isoindole axes (dashed yellow and green lines) for a single molecule and a molecule in a "++" dimer. Filled purple curves correspond to the difference between solid yellow and green curves. (b), (c) Constant-height, Laplace-filtered ncAFM images [tip approached 40 pm with respect to STM setpoint $V_b$ = 0.02 V, $I_t$ = 5 pA on bare Ag(100); CO-functionalized tip] for single molecule and "++" dimer ($d_{c-c}$ = 2.03 nm). Insets: ncAFM apparent height profiles along isoindole-

isoindole axes. (d), (e) DFT-calculated relaxed adsorption geometries for single molecule and "++" dimer ($d_{c-c}$ = 1.98 nm; yellow: Mg, black: C, blue: N, white: H, grey: Ag; isoindole unit is circled in red). Only top Ag layer is shown for clarity. (f) STM topographic asymmetry parameter $\alpha_{STM} = \frac{|(z_1+z_3)-(z_2+z_4)|}{(z_1+z_3+z_2+z_4)}$, for single molecule (blue) and molecule in "++"dimer (orange; $d_{c-c}$ = 2.1 nm), as a function of STM bias voltage, where $z_j$ ($j$ = 1, …, 4) is the maximum STM apparent height for each of the four isoindole groups, and $z_1 > z_3 > z_2 > z_4$ [see bottom right MgPc in (a) for labelling]. Error bars were determined by calculating extremal values of $\alpha_{STM}$ : $\alpha_{STM,max} = \frac{z_1-z_4}{z_1+z_4}$; $\alpha_{STM,min} = \frac{z_3-z_2}{z_3+z_2}$. Scale bars: 1 nm.

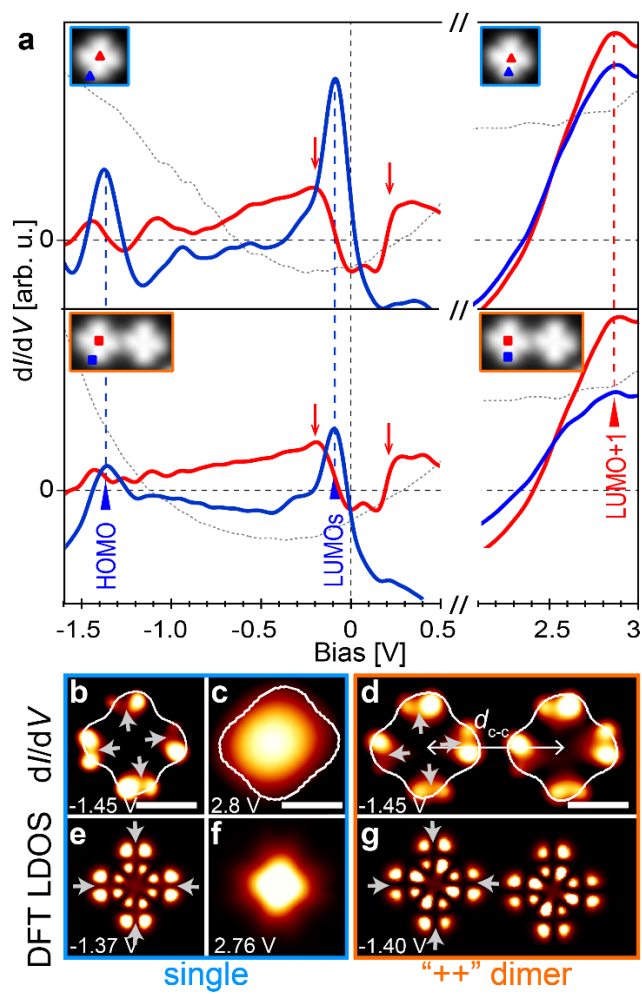

**Figure 2. Electronic structure of single MgPc and MgPc of a "++" dimer on Ag(100).** (a) d$I$/d$V$ spectra acquired on the phthalocyanine (Pc) ligand (blue) and Mg center (red) for single MgPc (top panel) and "++" dimer (bottom). d$I$/d$V$ spectra were background subtracted using the d$I$/d$V$ curve on bare Ag (dashed grey curve). Setpoints: $V_b$ = -2 V, $I_t$ = 3 nA for data from -1.6 to 0.5 V; $V_b$ = 3 V, $I_t$ = 50 pA for data from 2.1 to 3 V. Blue and red triangular ticks indicate features related to the gas-phase HOMO, the two-fold degenerate gas-phase LUMOs [partially filled here on Ag(100)] and higher energy unoccupied states of gas-phase MgPc. Red arrows indicate near-Fermi features related to molecular vibrational modes. (b), (d) d$I$/d$V$ maps at $V_b$ = -1.45 V for single MgPc and "++" dimer ($d_{c-c}$ = 2.27 nm; molecular STM topographic contour at $V_b$ = -2.5 V, $I_t$ = 100 pA superimposed in white), and (c) at $V_b$ = 2.8 V for single MgPc (acquisition details in Methods; molecular STM topographic contour at $V_b$ = 2.8 V, $I_t$ = 250 pA superimposed in white). (e) - (g) DFT-calculated d$I$/d$V$ maps for single MgPc and "++" dimer ($d_{c-c}$ = 1.98 nm) (see Methods for computational details). Grey arrows indicate the positions of the molecular orbital nodal planes normal to the Ag surface. Scale bars: 1 nm.

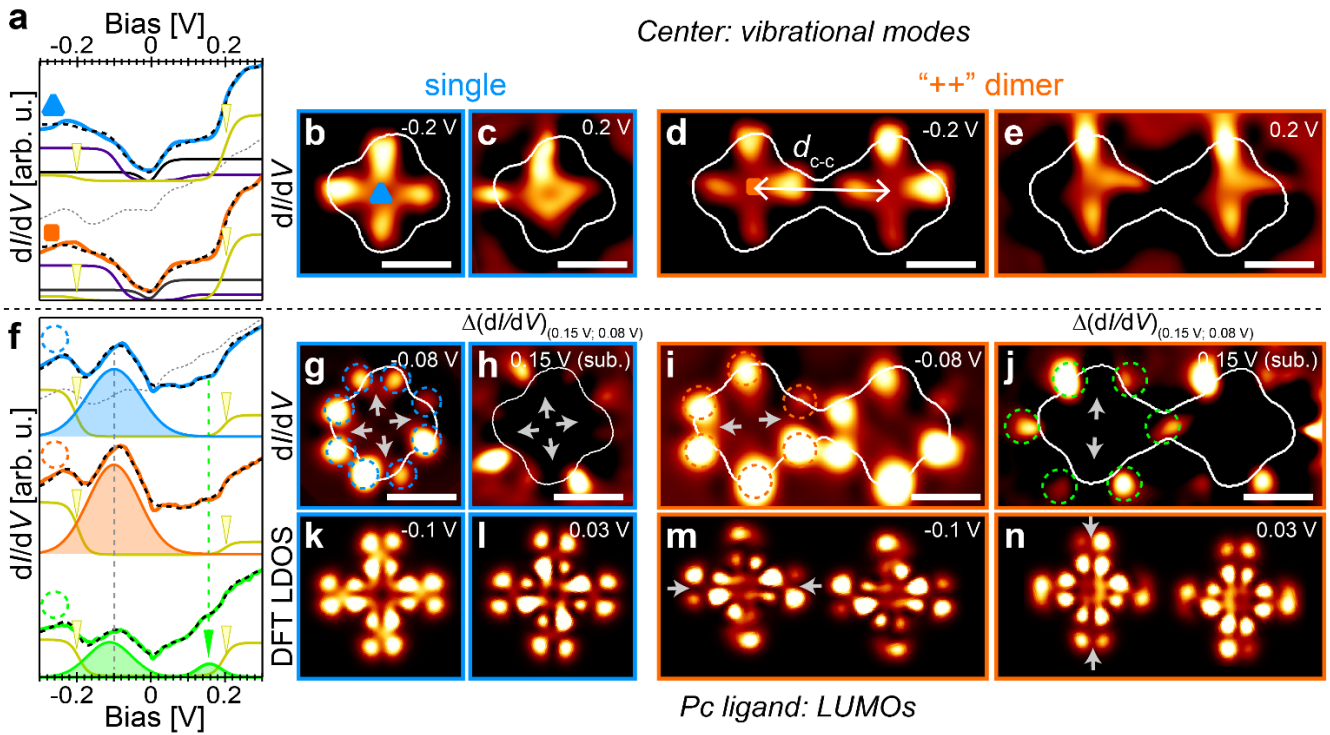

**Figure 3. Near-Fermi electronic structure: spatial symmetry breaking and degeneracy lifting of LUMOs.** (a) d$I$/d$V$ point spectra on Mg center for single MgPc (blue) and MgPc "++" dimer (orange; $d_{c-c}$ = 2.0 nm). Spectra were fit (black dashed curve) with a sum of Fermi-Dirac distributions (solid yellow, black, purple curves) and of the attenuated reference bare Ag(100) spectrum (SI section S6 for details). (b)-(e) d$I$/d$V$ maps at $V_b$ = ± 0.2 V for single MgPc (blue) and "++" dimer (orange). (f) d$I$/d$V$ point spectra (solid blue, orange, green curves) averaged over peripheral ligand regions indicated by dashed circles in (g), (i) and (j). Spectra were fit (black dashed curve) with a sum of Fermi-Dirac distributions (yellow curve), a Gaussian peak (filled curves) and an attenuated reference bare Ag(100) spectrum (grey dashed curve). (a) and (f) were acquired with the same tip and parameters (tip was approached 100 pm towards the sample, with respect to the STM setpoint $V_b$ = -2.5 V, $I_t$ = 400 pA). (g) - (j) d$I$/d$V$ maps at $V_b$ = -0.08 and 0.15 V for single MgPc and "++" dimer. Maps acquired at 0.15 V (h, j) were subtracted by maps acquired at 0.08 V (i.e. $\Delta(dI/dV)_{(0.15\ V;\ 0.08\ V)} = (dI/dV)_{0.15\ V} - (dI/dV)_{0.08\ V}$), to minimize inelastic contributions of molecular vibrational modes and emphasize elastic features of LUMOs. Grey arrows indicate orbital nodal planes. (k) - (n) Corresponding DFT-calculated differential conductance maps (Methods). White solid outlines indicate molecule STM topographic contour ($V_b$ = -2.5 V, $I_t$ = 100 pA). Scale bars: 1 nm.

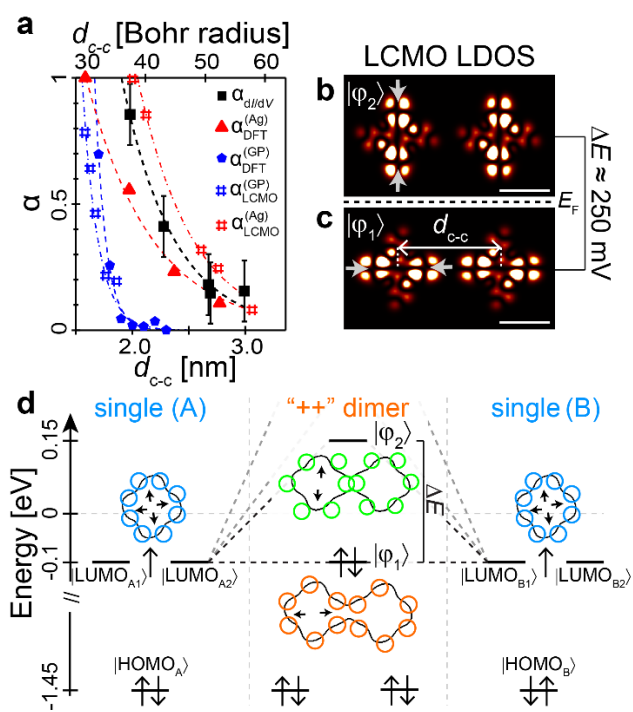

**Figure 4. Long-range molecule-molecule hybridization mediated by Ag(100) substrate.** (a) Spectroscopic asymmetry parameters $\alpha$ for MgPc in "++" dimer, as a function of intermolecular center-center distance $d_{c\text{-}c}$. Black squares: experimental $\alpha_{dI/dV}$ retrieved from d$I$/d$V$ maps at $V_b$ = -0.08 V (see SI section S10 and S11); error bars correspond to $\alpha_{dI/dV}$ calculated for a single isolated MgPc (which differs slightly from zero due to tip asymmetries; see SI). Red triangles (blue pentagons): $\alpha_{\text{DFT}}^{(\text{Ag})}$ ($\alpha_{\text{DFT}}^{(\text{GP})}$) extracted from DFT-simulated d$I$/d$V$ maps of "++" dimer on Ag(100) (in gas-phase, respectively); see Methods and SI Figure S19. Red (blue) hashes: $\alpha_{\text{LCMO}}^{(\text{Ag})}$ ($\alpha_{\text{LCMO}}^{(\text{GP})}$) extracted from Linear Combination of Molecular Orbitals (LCMO) model for "++" dimer on Ag(100) (in gas-phase, respectively). Dashed curves: decaying exponential fits. (b), (c) Calculated differential conductance maps (see Methods) corresponding to the two lowest energy hybrid orbitals $\varphi_1$ and $\varphi_2$ generated with our LCMO model (see SI section S14) for a "++" dimer ($d_{c\text{-}c} \cong$ 2.0 nm). Grey arrows indicate nodal planes. Scale bars: 1 nm. (d) Energy level diagram for single MgPc's $A$ and $B$ (left and right), each with doubly degenerate, partially occupied LUMOs, and a "++" dimer (center) with non-degenerate, occupied and unoccupied hybrid orbitals. Blue, green and orange contours represent spatial distributions of experimental d$I$/d$V$

maps in Figure 3g, i, j, respectively. Solid black curves: corresponding STM molecular contours ($V_b$ = -2.5 V, $I_t$ = 100 pA; Figure 3).

**Table of Contents:**

**Long-Range Surface-Assisted Molecule-Molecule Hybridization**

*Marina Castelli, Jack Hellerstedt, Cornelius Krull, Spiro Gicev, Lloyd C.L. Hollenberg, Muhammad Usman, Agustin Schiffrin*

On a silver surface, magnesium phthalocyanine molecules undergo a perturbation of their electronic structure as a result of an attractive interaction with their nearest-neighbors. Quantitative agreement with supporting theoretical modelling indicates that this interaction consists of multiple-nanometer-range intermolecular hybridization enabled by the underlying substrate. These observations offer new possibilities to control electronic properties for engineered nanomaterials.

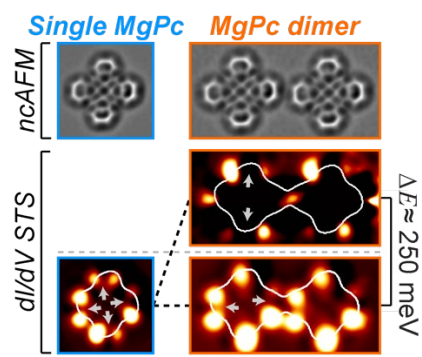

# Supporting Information

## Long-Range Surface-Assisted Molecule-Molecule Hybridization

*Marina Castelli, Jack Hellerstedt, Cornelius Krull, Spiro Gicev, Lloyd C.L. Hollenberg, Muhammad Usman*[*], *Agustin Schiffrin*[*]

Dr. M. Castelli, Dr. J. Hellerstedt, Dr. C. Krull, Dr. A. Schiffrin

School of Physics and Astronomy, Monash University, Clayton, Victoria 3800, Australia

Email: agustin.schiffrin@monash.edu

Dr. M. Castelli, Dr. C. Krull, Dr. A. Schiffrin

ARC Centre of Excellence in Future Low-Energy Electronics Technologies, Monash University, Clayton, Victoria 3800, Australia

S. Gicev, Dr. M. Usman, Prof. L.C.L Hollenberg

Centre for Quantum Computation and Communication Technology, School of Physics, The University of Melbourne, Parkville, 3010, Victoria, Australia

Email: muhammad.usman@unimelb.edu.au

Dr. M. Usman

School of Computing and Information Systems, Melbourne School of Engineering, The University of Melbourne, Parkville, 3010, Victoria, Australia

## Contents

## S1. Intramolecular conformation; registry of MgPc molecules on Ag(100)

We performed high resolution scanning tunneling microscopy (STM) and non-contact atomic force microscopy (ncAFM) with a tip functionalized with a carbon monoxide (CO) molecule, to determine the intramolecular conformation and registry of MgPc on Ag(100), for the "single" and "++" dimer cases (Figure S1; see also Figures 1b, c of main text). In both cases, the central Mg atom sits close to the top of a Ag(100) hollow site, with the intramolecular isoindole-isoindole axis (dashed black lines in Figures S1a, c) forming an angle of 29° ± 1° with respect to the [011] direction of the crystalline substrate. This is consistent with previous reports[1] of metal-phthalocyanines (M-Pc's) on Ag(100). Both CO-tip STM and ncAFM measurements show an intramolecular morphology that is very similar for both the single isolated molecule and a molecule in a "++" dimer. We conclude that MgPc is exposed to the same Ag environment in both cases, and that the adsorption configuration has no contribution to the observed break of four-fold rotational symmetry.

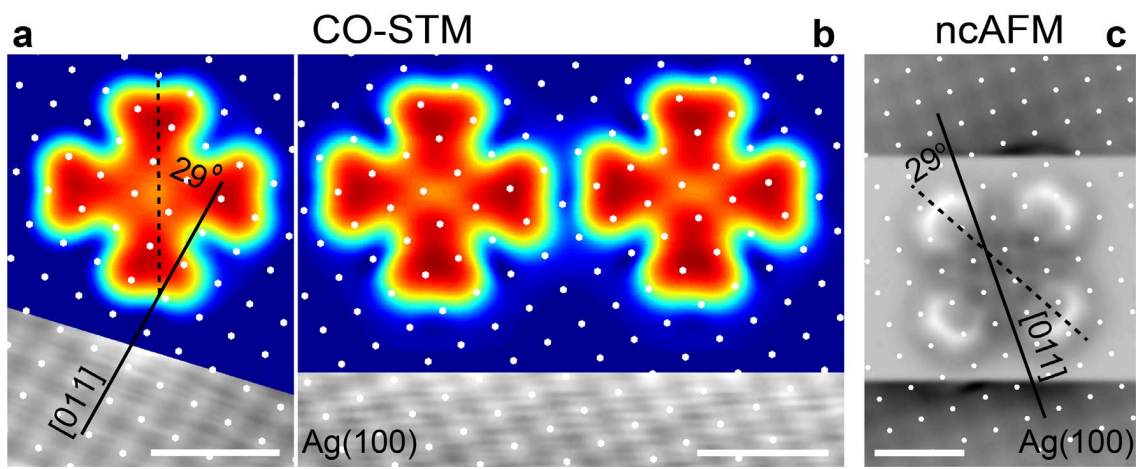

**Figure S1. Registry of MgPc on Ag(100) for the single and "++" dimer cases**. (a), (b) Constant-current CO-functionalized STM topographies of a single isolated MgPc and of a "++" dimer [$V_b$ = 0.05 V, $I_t$ = 10 pA on molecule; $V_b$ = 0.05 V, $I_t$ = 2 nA on Ag(100) to obtain the atomic resolution]. (c) Constant-height CO-tip ncAFM

image (Gaussian-smoothed) of a single MgPc. Tip height is defined with respect to the STM setpoint $V_b$ = 0.02 V, $I_t$ = 5 pA on the Ag(100) substrate. On the molecule, the tip was approached 40 pm from the reference setpoint towards the surface to obtain intramolecular resolution. On bare Ag, the tip was approached 300 pm towards the surface for resolving Ag atoms. Scale bars: 1 nm.

Figure S2 shows CO-tip ncAFM measurements of the frequency shift $\Delta f$ as a function of tip-sample distance $z$, acquired at different locations of the single MgPc and "++" dimer. The $z$ values associated with the $\Delta f$ minima – which can provide an indication of any possible difference in adsorption height[2] – are identical for both cases. In conjunction with constant-height CO-tip ncAFM imaging (insets of Figure S2; Figures 1b, c of main text), we conclude that the intramolecular morphology is identical for both single MgPc and "++" dimer. This is in agreement with DFT-calculated relaxed molecular structures (Figure S2b), showing flat molecular adsorption and no significant variation in molecular conformation between both cases (other than height differences of only a few pm, smaller than our experimental resolution). We therefore exclude structural distortion as being the cause of the observed reduced symmetry in Figure 1 of the main text. Note that both experimental data and DFT calculations reveal that the Mg atom is closer to the Ag surface than the plane defined by the Pc ligand.

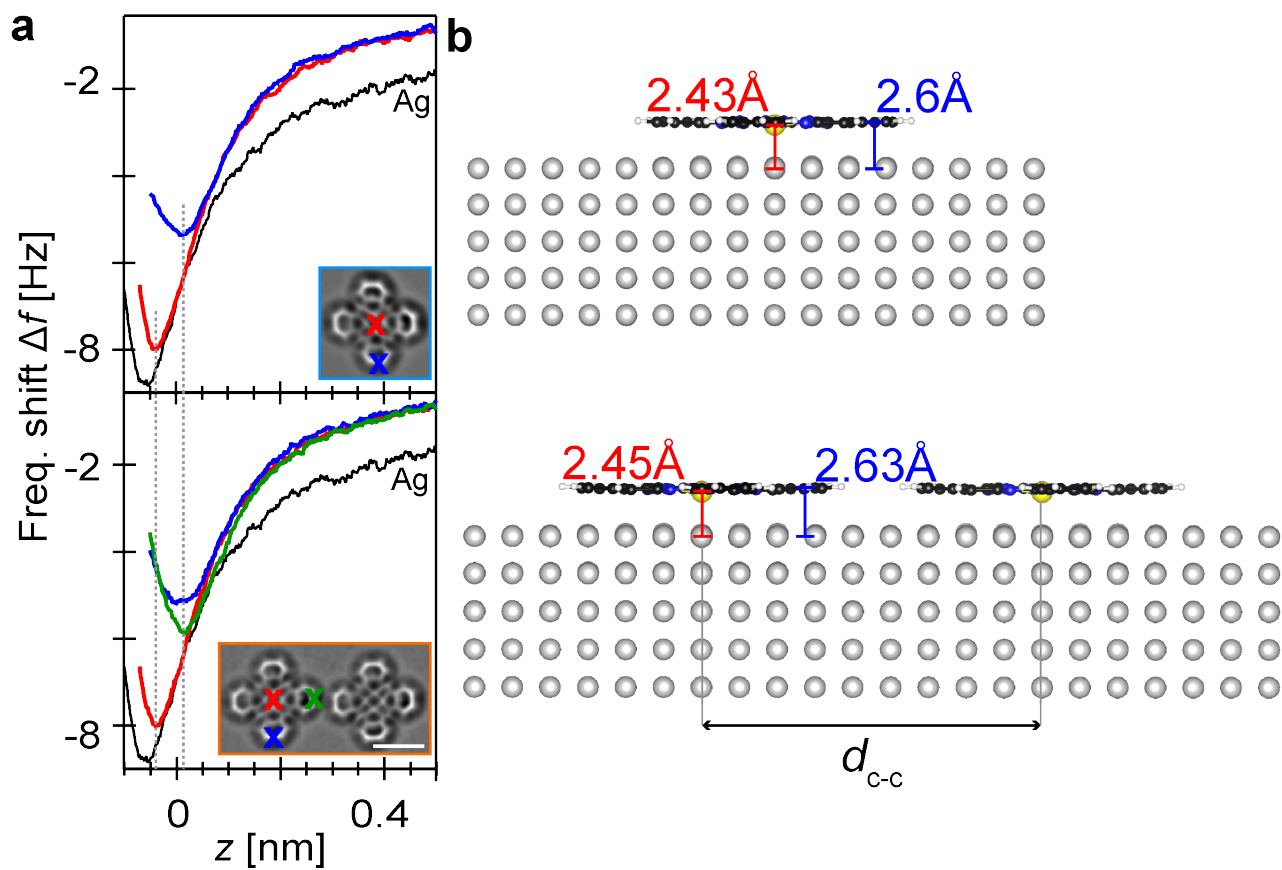

**Figure S2. Intramolecular conformation of single MgPc and "++" dimer: CO-tip ncAFM data and DFT calculations.** (a) Frequency shift $\Delta f$ as a function of tip-sample separation $z$, acquired at different locations [red: Mg center; blue, green: C-C bond at isoindole group extremity; black: bare Ag(100)] of the single isolated MgPc (top) and MgPc "++" dimer (bottom, $d_{c\text{-}c}$ = 1.98 nm). The tip-sample separation $z$ was determined with respect to the STM setpoint $V_b$ = 0.02 V, $I_t$ = 5 pA on bare Ag(100). Bias voltage $V_b$ was set to 1 mV during $\Delta f(z)$ acquisition. The $z$ values associated with the $\Delta f$ minima (dashed vertical lines) are identical for the single MgPc and "++" dimer, indicating the same intramolecular conformation in both cases. The central Mg atom sits ~20 pm closer to the surface than the isoindole groups. Insets: constant-height ncAFM images. (b) Side view of DFT-simulated relaxed structure of single MgPc and "++" dimer, showing similar intramolecular conformation for both cases (with molecular height differences of only a few pm), in agreement with experiments (grey: Ag; black: carbon; blue: nitrogen; white: hydrogen; yellow: magnesium).

## S2. Bias-dependent STM imaging of an MgPc "++" dimer

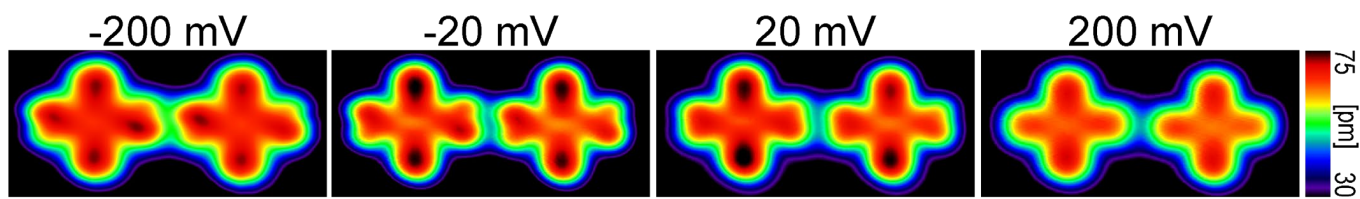

**Figure S3. Bias-dependent STM topography for "++" dimer.** Constant-current STM images of MgPc molecules in a "++" dimer ($I_t$ = 0.4 nA, $V_b$ = -200 mV, -20 mV, 20 mV, 200 mV respectively from left to right). Center-center distance: ~2.1 nm.

## S3. Deposition of carbon monoxide (CO) molecules onto Ag(100)

High resolution ncAFM imaging required functionalization of the tip with a carbon monoxide (CO) molecule[3]. To do so, we dosed CO controllably into the ultrahigh vacuum (UHV) chamber with the Ag(100) sample held at 4.6 K. At this temperature, CO adsorbs and remains immobile on Ag(100). However, it is not uncommon for CO molecules to interact with organic molecules and metal adatoms[4]. We performed STM imaging of large MgPc/Ag(100) sample areas upon CO deposition to check for potential contamination of the MgPc molecules (Figure S). Single molecules can be recognized by their characteristic cross-like shape; CO molecules appear as dark depressions, consistent with observations on Ag(111)[5].

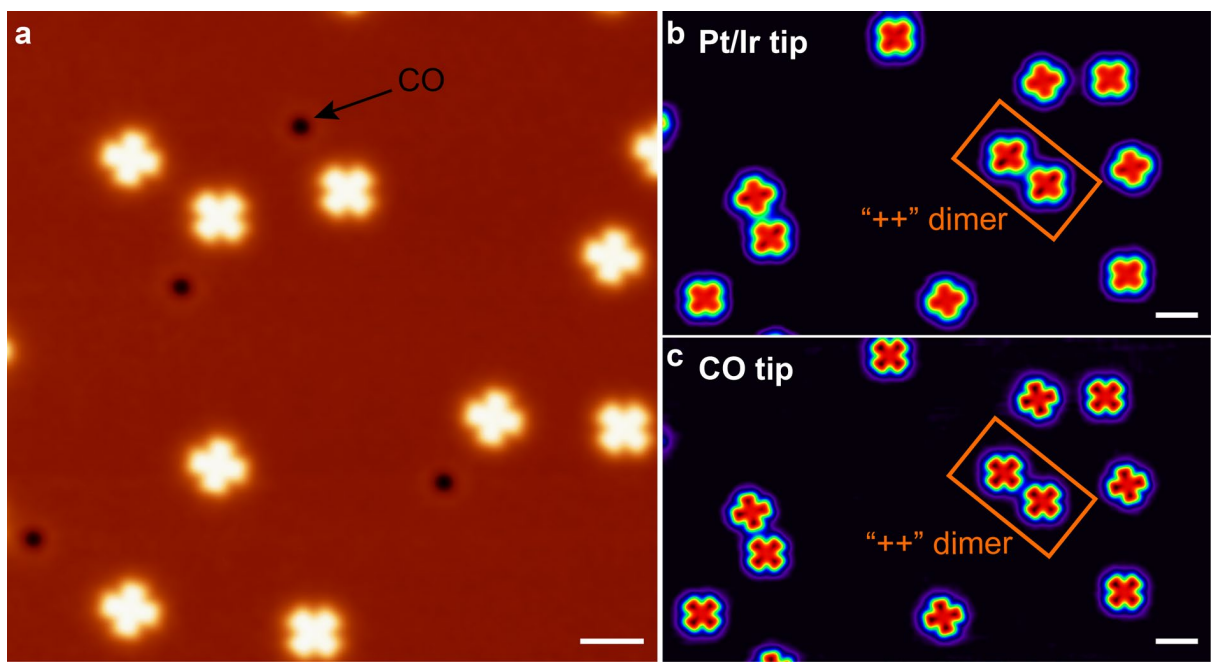

**Figure S4. CO and MgPc molecules on Ag(100).** (a) Constant-current STM image of MgPc molecules on Ag(100) after dosing CO ($V_b$ = 0.02 V, $I_t$ = 5 pA). CO does not seem to interact with MgPc. (b), (c) Constant-current STM images of MgPc on Ag(100) with a Pt/Ir tip (b) and a CO-functionalized tip (c) ($V_b$ = 0.02 V, $I_t$ = 5 pA). Scale bars: 2 nm.

Figures S4b, c correspond to STM images of MgPc on Ag(100), before and after functionalizing the tip with CO, respectively. Note that the topographic asymmetry of the "++" dimer visible with the metallic Pt/Ir tip vanishes with the CO-functionalized tip. Given the electronic nature of the observed break of symmetry, we attribute this to changes of the junction electronic properties induced by the CO termination[5a].

## S4. Long-range STM imaging of MgPc molecules on Ag(100)

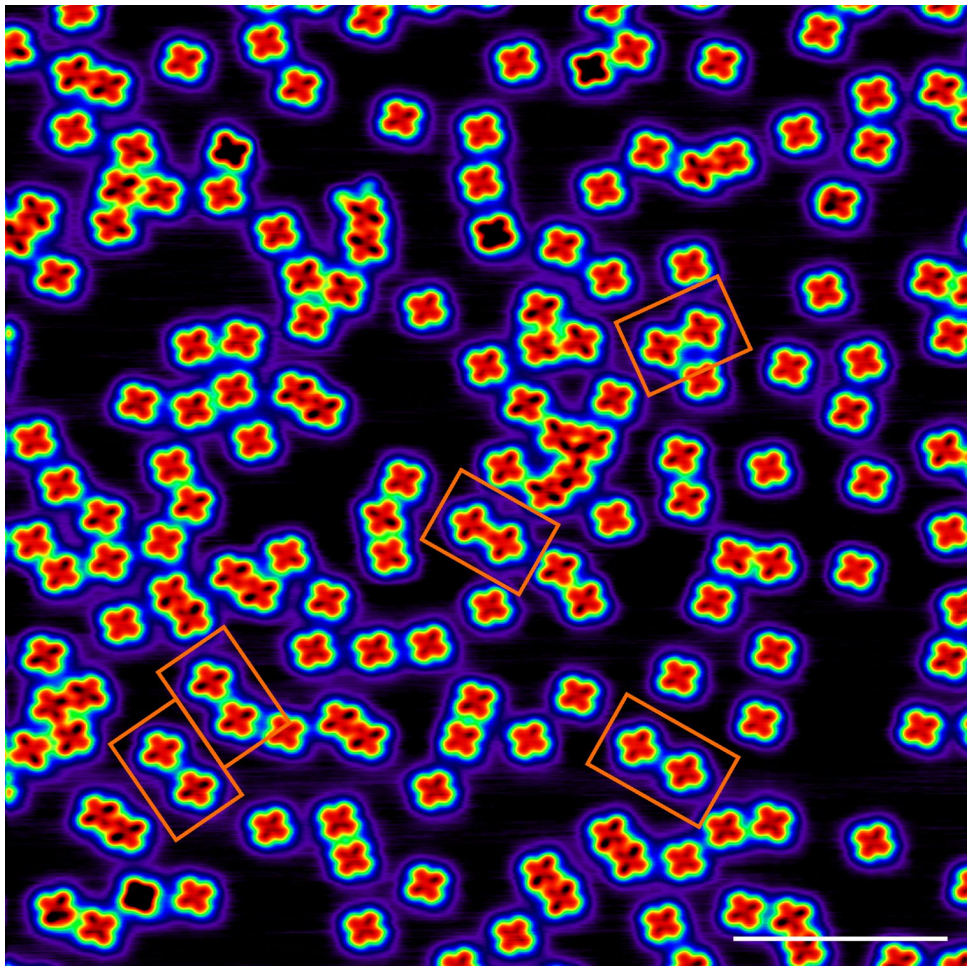

**Figure S5. Constant-current large-scale STM image of MgPc molecules on Ag(100).** Examples of "++" dimers with broken four-fold rotational symmetry are framed in orange ($I_t$ = 25 pA, $V_b$=100 mV). Scale bar: 10 nm.

## S5. Treatment of d$I$/d$V$ STS data

Within the bias voltage range that we considered, the local density of electronic states (LDOS) of bare Ag(100) does not yield any sharp or abrupt spectroscopic feature (e.g., no Shockley surface state). It is hence possible to mitigate, in a d$I$/d$V$ STS spectrum of interest (here, of MgPc for example), effects due to possible tip electronic states[6] or exponential tunneling transmission[7], by subtracting a reference d$I$/d$V$ spectrum taken on bare Ag(100) with the same parameters. Figure S6 shows an example of MgPc d$I$/d$V$ spectra on which such background subtraction procedure was performed. After subtracting the bare

Ag(100) background curve (dashed grey curves in Figures S6a, b), features related to intrinsic molecular electronic states (HOMO-1, HOMO, LUMOs indicated with blue ticks) become more prominent. It is important to note that this background subtraction can lead to negative differential conductance (NDC) due to attenuation of the spectroscopic signature of Ag(100) and the tip when tunneling through the molecule. This NDC has no physical meaning. It is important to consider values of background-subtracted d$I$/d$V$ spectra relatively.

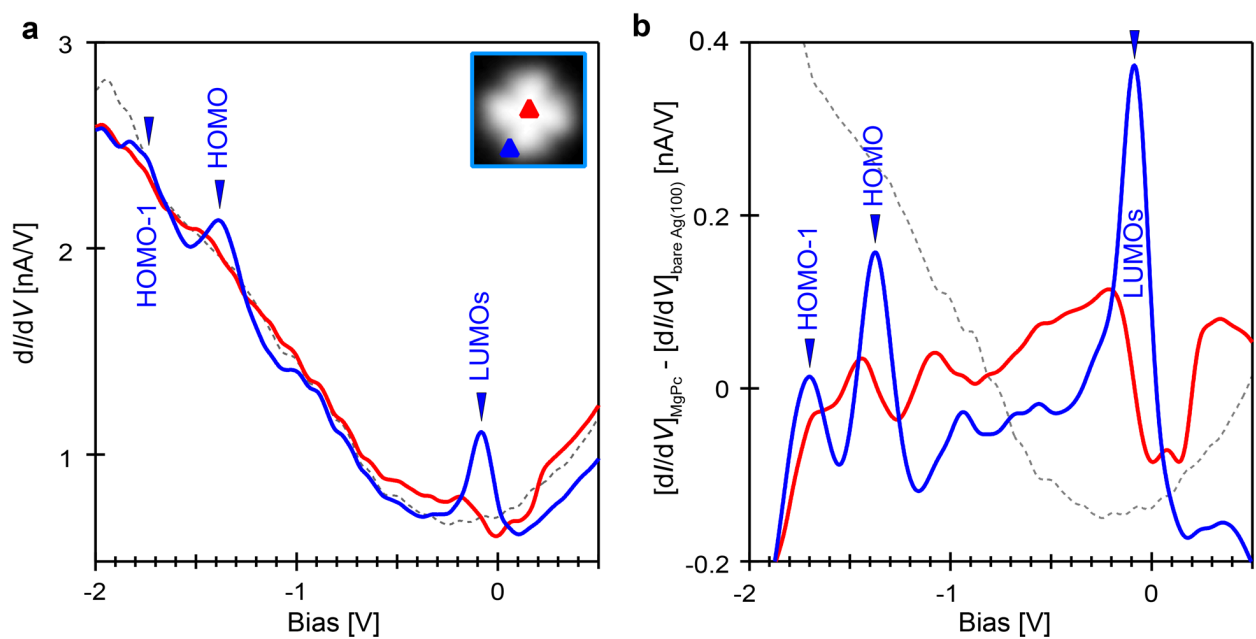

**Figure S6. d$I$/d$V$ STS of single MgPc: background subtraction.** (a) Numerical d$I$/d$V$ point spectra of bare Ag (dashed grey), and of single MgPc center (red) and Pc ligand (blue). STM setpoint: $V_b$ = -2 V, $I_t$ = 3 nA. (b) Same d$I$/d$V$ spectra after subtraction of bare Ag(100) spectrum (grey). Exponential transmission is reduced and molecular electronic states (HOMO-1, HOMO, LUMOs marked with blue ticks) are accentuated.

## S6.  Fitting of near-Fermi d$I$/d$V$ spectra

Figures 3a, f of the main text consist of d$I$/d$V$ spectra of a single MgPc molecule and a molecule in a "++" dimer, for small bias voltage absolute values (i.e., near the Fermi level). We fit each of these curves with the general function:

S8

$$f_{\text{fit}}(V_{\text{b}}) = a \cdot (dI/dV)_{\text{Ag}}(V_{\text{b}}) + \sum_{k=1}^{n_{\text{el}}} b_k \cdot g_k(V_{\text{b}}) + \sum_{l=1}^{n_{\text{vibr}}} \left[ c_l \cdot f_{\text{FD},l}(V_{\text{b}}) + d_l \cdot \left(1 - f_{\text{FD},l}(-V_{\text{b}})\right) \right]$$

where variable $V_{\text{b}}$ is the bias voltage, $(dI/dV)_{\text{Ag}}$ is a $dI/dV$ spectrum acquired on bare Ag(100) (i.e., the fit takes account of tip and bare substrate electronic features), $g_k(V_{\text{b}}) = \exp\left(-\frac{\left(V_{\text{b}} - V_k^{(\text{el})}\right)^2}{2\sigma_k^2}\right)$ are Gaussian functions accounting for $n_{\text{el}}$ molecular electronic states, $f_{\text{FD},l}(V_{\text{b}}) = \frac{1}{\exp\left(\left(V_{\text{b}} - V_l^{(\text{vibr})}\right)/k_{\text{B}}T\right) + 1}$ are Fermi-Dirac distributions [$c_l \cdot f_{\text{FD},l}(V_{\text{b}})$ and $d_l \cdot \left(1 - f_{\text{FD},l}(-V_{\text{b}})\right)$ ensure energy-symmetric Fermi-Dirac steps with respect to the Fermi level] accounting for $n_{\text{vibr}}$ molecular vibrational modes, and $a$, $b_k$, $c_l$, $d_l$, $V_k^{(\text{el})}$, $\sigma_k$, $V_l^{(\text{vibr})}$ and $T$ are fitting parameters.

a)    **Fitting of d$I$/d$V$ spectra on central Mg**

For the d$I$/d$V$ spectrum acquired at the Mg center of the MgPc molecule (Figures 3a in main text), we considered three different fitting scenarios A1-A3 [Figure S7; all fits include the background $(dI/dV)_{\text{Ag}}$ contribution, i.e., $a \neq 0$; errors correspond to 95% confidence interval of fit]:

A1. Fitting d$I$/d$V$ spectrum assuming one electronic resonance ($n_{\text{el}} = 1$; $V_1^{(\text{el})} = -0.150 \pm 0.001$ V) and one vibrational mode ($n_{\text{vibr}} = 1$; $V_1^{(\text{vibr})} = -0.198 \pm 0.001$ V; Figure S7a). The electronic resonance was motivated by the observation of the LUMOs-related d$I$/d$V$ feature on the peripheral Pc ligand at $V_{\text{b}} \cong -0.1$ V (Figure 2a of main text). The vibrational mode at ~ ±200 mV is consistent with previous works[8].

A2. Fitting d$I$/d$V$ spectrum assuming two vibrational modes ($n_{\text{vibr}} = 2$; $V_1^{(\text{vibr})} \cong -0.202 \pm 0.002$ V; $V_2^{(\text{vibr})} \cong -0.089 \pm 0.001$ V; Figure S7b), without the electronic resonance at $V_{\text{b}} \cong -0.1$ V ($n_{\text{el}} = 0$, since this feature is mainly located at the peripheral Pc ligand; see Figure 2a of main text).

A3. Fitting d$I$/d$V$ spectrum assuming three vibrational modes ($n_{vibr} = 3$; $V_1^{(vibr)} \cong -0.005 \pm 0.001$ V; $V_2^{(vibr)} \cong -0.089 \pm 0.001$ V; $V_3^{(vibr)} \cong -0.2 \pm 0.001$ V; Figure S7c), without the electronic resonance at $V_b \cong -0.1$ V ($n_{el} = 0$). The energies of these vibrational modes are consistent with previous works[1a, 8-9].

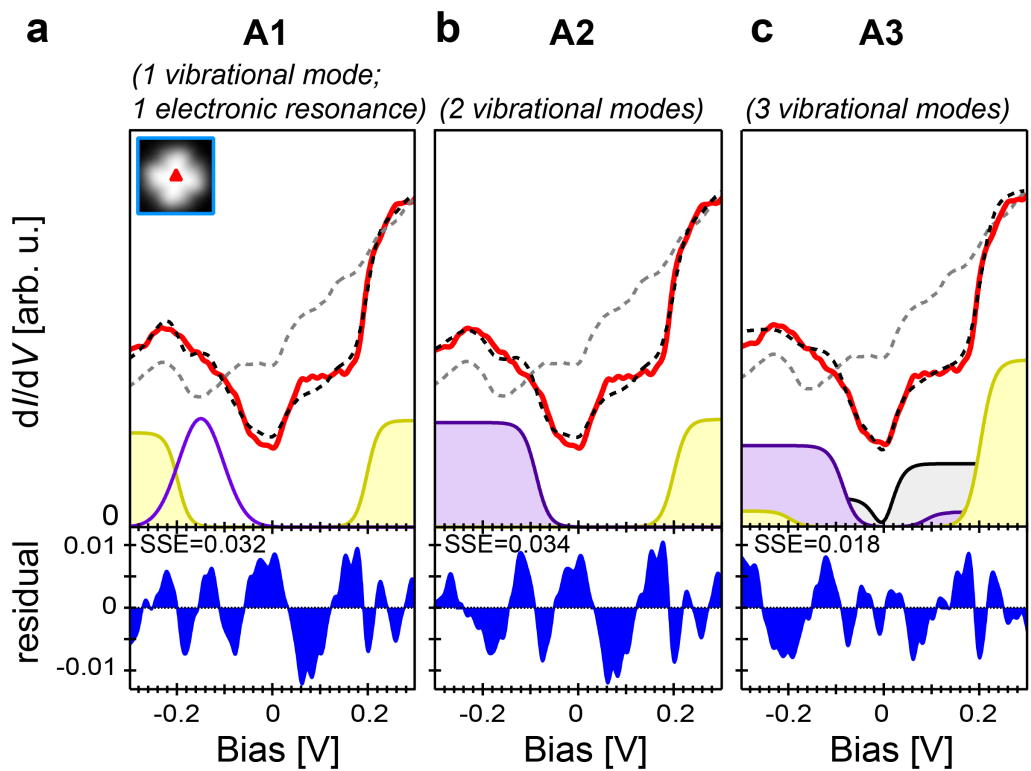

**Figure S7.** (a) Fitting of experimental d$I$/d$V$ spectrum (solid red curve) acquired at center of single MgPc (inset: STM image), according to scenario A1. Dashed black curve: total fit; solid purple: Gaussian function accounting for electronic resonance; filled yellow: Fermi-Dirac distributions accounting for vibrational mode; solid grey: Ag(100) reference spectrum; filled blue: residual difference between experimental data and fit. (b) Same as (a), for scenario A2. Filled purple and yellow curves: Fermi-Dirac distributions accounting for vibrational modes. (c) Same as (a), for scenario A3. Filled purple, yellow and grey curves: Fermi-Dirac distributions accounting for vibrational modes. The sum of squared estimate of errors (SSE) is significantly smaller for scenario A3. Curves vertically offset for clarity.

In comparison to fitting scenarios A1 and A2, scenario A3 results in the smallest sum of squared estimate of errors (SSE); see Figure S7c. We further tried to add a Gaussian peak at $V_b \cong -0.1$ V to fitting scenario A3 (to potentially account for the electronic resonance associated with the LUMOs; see fitting scenario

A1 in Figure S7a); however, this increases the SSE and results in a worse fit. Adding more vibrational modes also failed to improve the fit. For these reasons we decided to model the Mg center d$I$/d$V$ STS spectra according to scenario A3, by taking into account three vibrational modes and the attenuated background signal given by the the tip and Ag(100).

**b)    Fitting of d$I$/d$V$ spectra at peripheral Pc ligand**

Similar to the Mg center above, we fit the d$I$/d$V$ spectra acquired at the peripheral Pc ligand of a single MgPc and of an MgPc "++" dimer (Figure 3f in main text), by considering two different fitting scenarios B1 and B2. The fitting curves include the background $(dI/dV)_{Ag}$ contribution, i.e., $a \neq 0$.

B1. Fitting d$I$/d$V$ spectrum assuming one electronic resonance ($n_{el} = 1$) and one vibrational mode ($n_{vibr} = 1$).

B2. Fitting d$I$/d$V$ spectrum assuming two electronic resonances ($n_{el} = 2$) and one vibrational mode ($n_{vibr} = 1$).

For the single isolated MgPc, applying scenario B1 to fit the peripheral Pc average d$I$/d$V$ spectrum (light blue curve in Figure 3f of main text) resulted in a good fit ($V_1^{(el)} = -0.097 \pm 0.001$ V; $V_1^{(vibr)} = -0.202 \pm 0.001$ V). We attribute the below-Fermi electronic resonance to the degenerate, partially populated LUMOs (see d$I$/d$V$ map in Figure 3g of main text).

The motivation of B2 (with two electronic resonances) stems from the fact that d$I$/d$V$ mapping of the MgPc "++" dimer hints at two electronic resonances, one below ($V_b = -0.098 \pm 0.001$ V; Figure 3i of main text) and one above Fermi ($V_b = 0.157 \pm 0.002$ V; see Figure 3j of main text where the d$I$/dV map at $V_b = 150$ mV was subtracted by the d$I$/d$V$ map at $V_b = 80$ mV to minimize features related to molecular vibrational modes; see Figure 3a-e of main text). The average d$I$/d$V$ spectrum (orange curve in Figure 3f of main text), taken at areas of the "++" dimer where the d$I$/d$V$ signal at $V_b \cong -100$ mV is large (dashed

S11

orange circles in Figure 3i of main text), was also fit well with scenario B1 ($V_1^{(el)} = -0.098 \pm 0.001$ V; $V_1^{(vibr)} \cong -0.198 \pm 0.001$ V). We attribute the below-Fermi electronic resonance (associated with the Gaussian peak) to the degeneracy-lifted, partially populated low-energy LUMO.

The average d$I$/d$V$ spectrum (green curve in Figure 3f of main text and Figure S8), at areas of the "++" dimer where the d$I$/d$V$ signal at $V_b \cong 150$ mV is large (dashed green circles in Figure 3j of main text), was not fit well by scenario B1, as emphasized by the large non-random residual between $V_b \cong 0$ and ~200 mV in Figure S8a. This spectrum was fit better by scenario B2 ($V_1^{(el)} = -0.11 \pm 0.002$ V; $V_2^{(el)} = 0.157 \pm 0.002$ V; $V_1^{(vibr)} = -0.196 \pm 0.001$ V), with a significant reduction of the residual and of the SSE (Figure S8b). We attribute the above-Fermi, second electronic resonance to a degeneracy-lifted, empty LUMO.

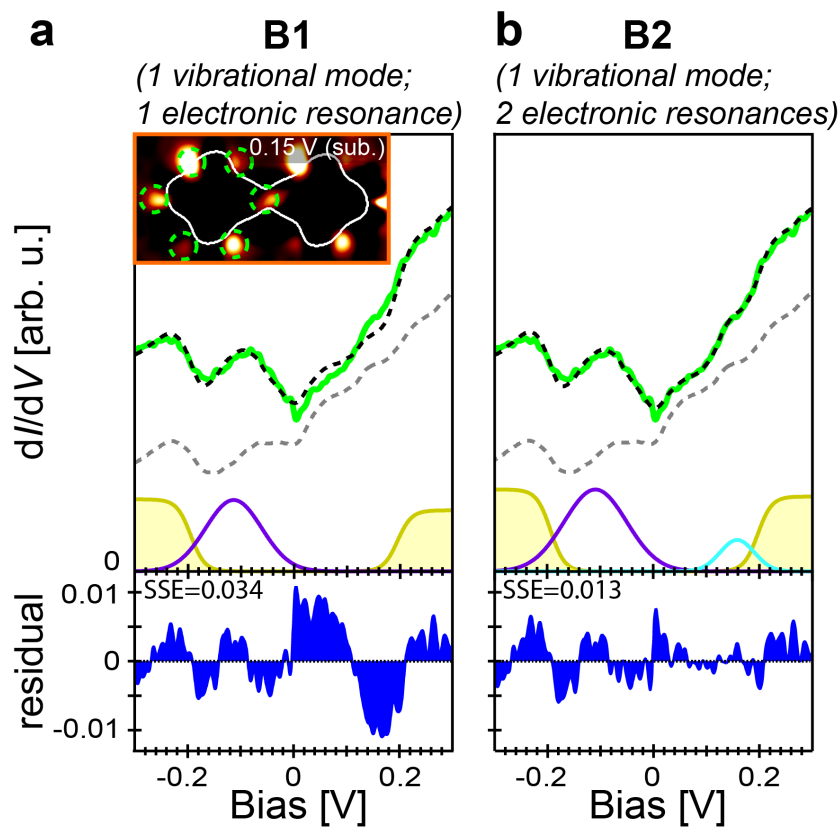

**Figure S8.** (a) Fitting of experimental average d$I$/d$V$ spectrum (solid green curve) acquired at peripheral Pc ligand of MgPc in "++" dimer (inset: d$I$/d$V$ difference map in Figure 3j of main text; dashed green circles indicate areas where spectra were acquired), according to scenario B1. Dashed black curve: total fit; solid purple: Gaussian function accounting for electronic resonance; filled yellow: Fermi-Dirac distributions accounting for vibrational mode; dashed grey: Ag(100) reference spectrum; filled blue: residual difference between experimental data and fit. (b) Same as (a), according to scenario B2. Solid cyan curve: Gaussian function accounting for 2$^{nd}$ electronic resonance. The SSE is significantly smaller for scenario B2. Curves vertically offset for clarity.

## S7. d$I$/d$V$ map acquisition: multipass (MP) technique

It is standard practice to acquire spatially resolved d$I$/d$V$ STS maps using a lock-in amplifier with an active feedback loop maintaining the tunneling current constant[10]. With this method, variations in the spatial distribution of the d$I$/d$V$ maps taken at different bias voltages can not only result from energy-dependent spectroscopic features [i.e., local density of electronic states (LDOS), vibrational modes], but also from bias-dependent changes in the STM apparent topography. This convolution between topography and spectroscopic properties can lead to erroneous conclusions regarding the physical properties of the system of interest. In our case, the limitation of this approach comes from the bias-dependent difference observed in STM imaging between a single isolated MgPc with apparent four-fold rotational symmetry, and an MgPc in a "++" dimer with reduced two-fold rotational symmetry (Figure 1 in main text).

We avoided this problem by acquiring d$I$/d$V$ maps with a multipass (MP) method, similar to that introduced by Moreno *et al.*[11] This MP method consists of a double sweeping scan: for each line of the scan, a first sweep is performed in constant-current STM mode at a specific setpoint, recording the surface topography (e.g., see Figure S9 below); a second sweep is then performed with the tip following the topography recorded during the first sweep, with the signal of interest (here, the d$I$/d$V$ signal provided by the lock-in amplifier) being recorded simultaneously. That is, during a MP scan of a sample region of interest, two datasets are recorded: the constant-current STM topography at the specific setpoint, and the d$I$/d$V$ signal acquired while the tip follows such STM topography.

In our case, for MgPc on Ag(100), STM topography during a MP scan was acquired at a bias voltage $V_b$ = -2.5 V. At this bias voltage, the STM topographies of both single isolated MgPc and MgPc in a "++" dimer (even in the case where the intermolecular distance is minimum; see Figure S9 below) are identical, with four-fold rotational symmetry. That is, d$I$/d$V$ maps acquired with this MP technique, with $V_b$ = -2.5 V during the constant-current STM topography acquisition, for single MgPc and for "++" dimer, are not subject to variations due to possible bias-dependent differences in STM apparent topography; any spatial variations observed in our MP-acquired d$I$/d$V$ maps are due to changes in spectroscopic properties (i.e., LDOS, molecular vibrational modes).

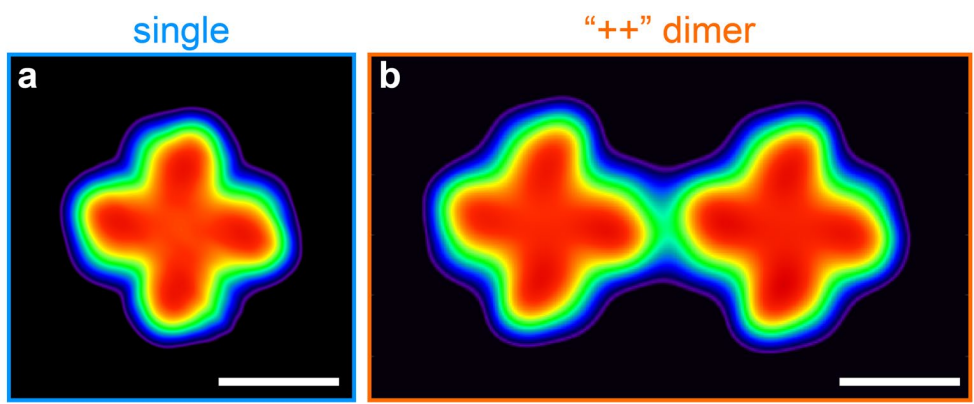

**Figure S9.** Constant-current STM image of (a) a single Mg and (b) an MgPc "++" dimer with minimum Mg-Mg intermolecular distance ($d_{c-c}$ = 2 nm), obtained during multipass (MP) d$I$/d$V$ map acquisition ($I_t$ = 100 pA, $V_b$ = -2.5 V). At this bias voltage of -2.5 V, all MgPc molecules (isolated and "++" dimer) appear with four-fold rotational symmetry. Scale bar: 1 nm.

Given the two-sweep approach, data obtained with the MP technique are highly sensitive to vibrations and tip instabilities. Figure S10a shows an example of an MP-acquired d$I$/d$V$ map. Background noise was attenuated by Gaussian smoothing, but dark and bright stripes remain on the left of the MgPc "++" dimer (white arrows). To correct for these artefacts, without losing d$I$/d$V$ spatial contrast related to the molecules, we sequentially applied the following filtering operations: (i) dilation [Matlab `imdilate()` function with a 0.02 nm radius disk structuring element; Figure S10b]; (ii) erosion [Matlab `imerode()` function with a 0.04 nm radius disk structuring element; Figure S10c]. This sequential filtering process allowed for removal of noise artefacts without changing the molecular d$I$/d$V$ spatial contrast.

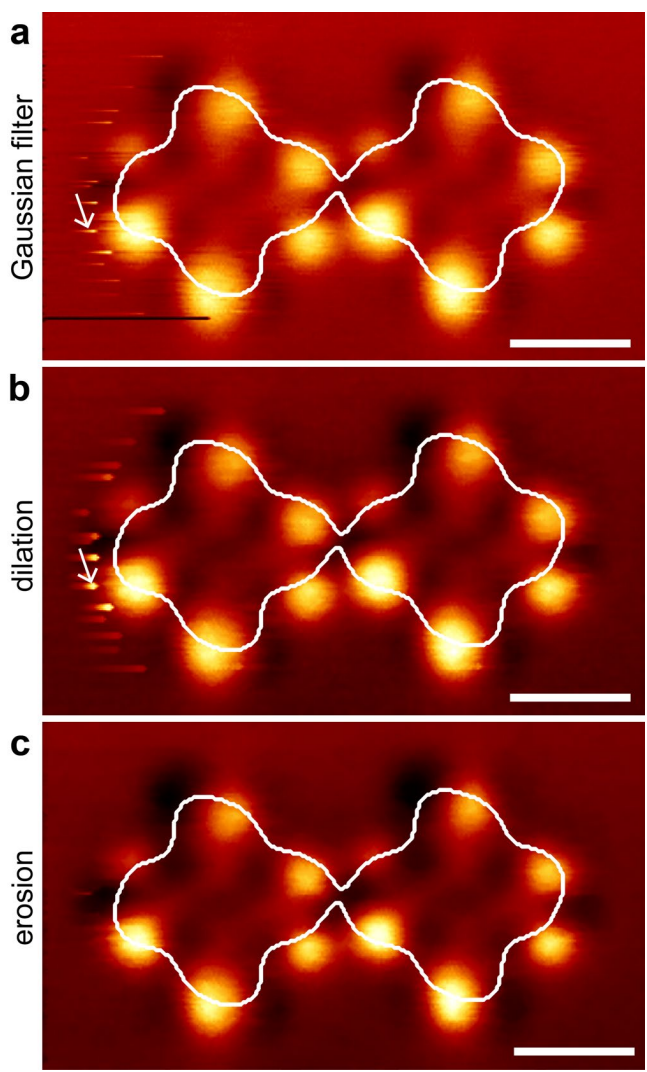

**Figure S10.** (a) MP-acquired d$I$/d$V$ map of "++" dimer ($d_{c-c}$ = 2 nm; STM topography: $I_t$ = 100 pA, $V_b$ = -2.5 V; d$I$/d$V$: $V_b$ = -0.08 V with tip approached 150 pm towards the surface) after sequential Gaussian filtering (10 consecutive convolutions with 0.01 nm FWHM Gaussian function), (b) dilation filtering (with 0.02 nm radius disk structuring element) and (c) erosion filtering (with 0.04 nm radius disk structuring element). Scale bar: 1nm.

Figure 3 in the main text shows MP-acquired d$I$/d$V$ maps for bias voltages $V_b$ between -0.2 and +0.2 V. Maps between -0.2 V and the Fermi level (Figure 3 in the main text; Figure S11a for $V_b$ = -0.08 V and b -0.04 V) are dominated by features of the partially populated LUMOs. For $V_b$ between the Fermi level and +0.2 V, the contrast of the d$I$/d$V$ maps is significantly different, with dominant intensity at the isoindole groups related to molecular vibrational modes (Figure 3 in the main text; Figure S11c to e for $V_b$ = 0.04 to 0.15 V).

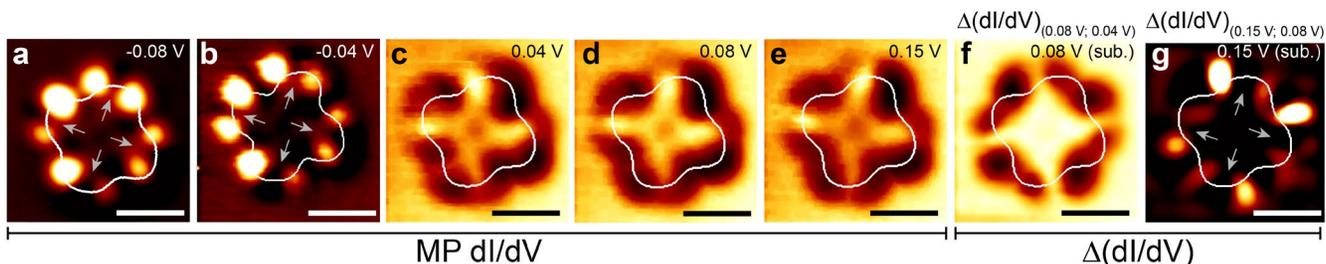

**Figure S11.** MP-acquired d$I$/d$V$ maps for a single MgPc on Ag(100), for (a) $V_b$ = -0.08 V, (b) -0.04 V, (c) 0.04, (d) 0.08 V and (d) 0.15 V (STM topography molecular contour in white: $I_t$ = 100 pA $V_b$ = -2.5 V). Difference d$I$/d$V$ maps [$\Delta(\mathbf{dI/dV})$], resulting from the subtraction of the MP-acquired d$I$/d$V$ map at (f) 0.08 V with that at 0.04 V [$\Delta(\mathbf{dI/dV})_{(0.08\ V;\ 0.04\ V)} = (\mathbf{dI/dV})_{0.08\ V} - (\mathbf{dI/dV})_{0.04\ V}$], and (g) from the subtraction of the MP-acquired d$I$/d$V$ map at 0.15 V with that at 0.08 V [$\Delta(\mathbf{dI/dV})_{(0.15\ V;\ 0.08\ V)} = (\mathbf{dI/dV})_{0.15\ V} - (\mathbf{dI/dV})_{0.08\ V}$]. These difference d$I$/d$V$ maps [$\Delta(\mathbf{dI/dV})$] were Gaussian and Laplacian (i.e., edge detection) filtered to enhance contrast and emphasize the orbital nodal planes in (g) (indicated with grey arrows). Scale bars: 1 nm.

In order to disentangle in d$I$/d$V$ STS contributions from local density of electronic states (LDOS) and those from molecular vibrational modes (e.g., see d$I$/d$V$ spectra in Figures 3a, f of main text), we considered difference d$I$/d$V$ maps, $\Delta(\mathrm{d}I/\mathrm{d}V)_{(V_1;\ V_2)} = (\mathrm{d}I/\mathrm{d}V)_{V_1} - (\mathrm{d}I/\mathrm{d}V)_{V_2}$, consisting of the subtraction of an MP-acquired d$I$/d$V$ map at a bias voltage $V_1$ with that at a bias voltage $V_2$ (Figure S11f, g). The difference map $\Delta(\mathrm{d}I/\mathrm{d}V)_{(0.08\ V, 0.04\ V)}$ in Figure S11f is dominated by features at the molecule center given by molecular vibrational modes (see fitting of d$I$/d$V$ spectra in section S6 above, and Figure S12). The d$I$/d$V$ map at 0.04 V (Figure S11c) shows a small contribution at the Pc periphery attributed to the partially occupied LUMOs; this manifests as the negative (dark) contrast at the isoindole edges in the $\Delta(\mathrm{d}I/\mathrm{d}V)_{(0.08\ V;\ 0.04\ V)}$ map (Figure S11f). The difference map $\Delta(\mathrm{d}I/\mathrm{d}V)_{(0.15\ V;\ 0.08\ V)}$ between d$I$/d$V$ maps at 0.15 and 0.08 V (Figure S11g) shows contrast at the peripheral Pc ligand, resembling the spatial distribution of the d$I$/d$V$ maps at -0.08 and -0.04 V (Figures S11a, b) related to the partially populated LUMOs (see grey arrows indicating LUMOs nodal planes). We rationalize this difference map $\Delta(\mathrm{d}I/\mathrm{d}V)_{(0.15\ V;\ 0.08\ V)}$ as a consequence of the LUMO+$U$, observed more prominently at a bias voltage of ~500 mV (see section S9 below), 'leaking' to lower energies due to energy broadening given by hybridization with the substrate.

## S8. d$I$/d$V$ STS of single MgPc: near-Fermi electronic and vibrational structure

In the main text, we show d$I$/d$V$ STS maps acquired either with the MP approach (Figures 2b, d, 3b-e, 3g-j; see section S7 above) or in constant-current mode with a lock-in amplifier (see Figure 2c and Methods in main text). The content of these maps sensitively depends on the choice of acquisition technique and experimental parameters. In Figures S12b-n, we compare near-Fermi d$I$/d$V$ maps acquired by the MP approach and by pixel-by-pixel numerical derivation of averaged $I(V)$ curves (that is, similar to d$I$/d$V$ spectra in Figure S12a and Figures 2a, 3a and 3f of main text; see Methods in main text).

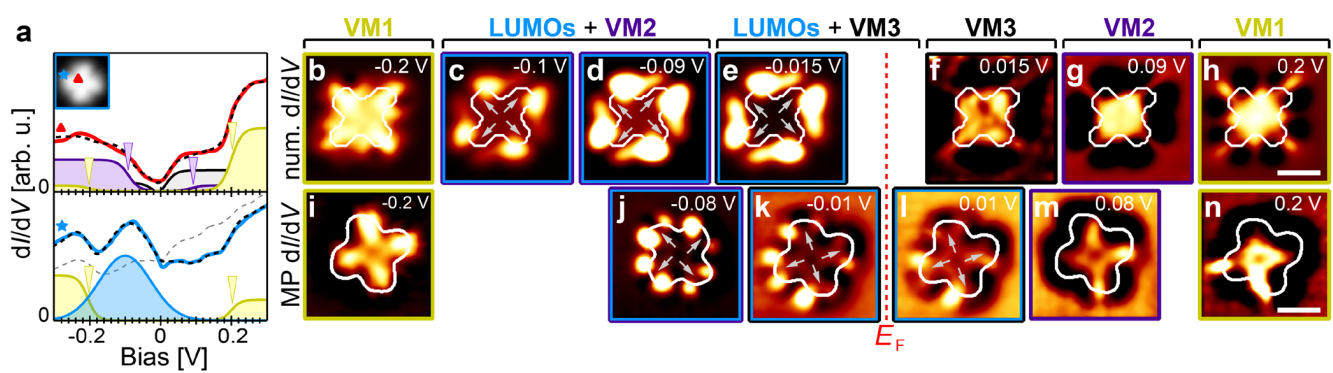

**Figure S12. Near-Fermi d$I$/d$V$ STS of single MgPc: electronic and vibrational structure.** (a) d$I$/d$V$ spectra at Mg center (red curve) and Pc ligand (blue). Tip approached 100 pm towards sample with respect to STM setpoint $V_b$ = -2.5 V, $I_t$ = 400 pA. Dashed black curves: total fits; dashed grey: Ag(100) reference spectrum; solid filled yellow, purple and black: fitting Fermi-Dirac distributions accounting for vibrational modes; solid filled blue: fitting Gaussian peak associated with LUMOs (see above and Figure 3a of main text). Inset: STM image of single MgPc. (b)-(h) d$I$/d$V$ maps obtained via pixel-by-pixel numerical derivation of averaged $I(V)$ curves (setpoint: $V_b$ = 0.015 V; $I_t$ = 20 pA; see Methods in main text). (i)-(n) Corresponding d$I$/d$V$ maps acquired via the MP technique (STM topography: $I_t$ = 400 pA, $V_b$ = -2.5 V; d$I$/d$V$: tip approached 150 pm towards the surface, details in section S7). Corresponding $V_b$ labelled on each panel. Frame colors indicate dominant contributions to the d$I$/d$V$ map from the corresponding Gaussian peak and Fermi-Dirac functions in (a). Maps show strong contributions from molecular vibrational modes [labelled VM1, VM2, VM3] at center of molecule. Between ~ -0.1 V and the Fermi level, contributions from LUMOs dominate, with strong d$I$/d$V$ signal at peripheral Pc ligand (grey arrows indicate LUMOs nodal planes). Scale bar: 1 nm.

Fitting of the Mg center d$I$/d$V$ spectra in Figure 3a of main text, and Figure S12a, includes Fermi-Dirac distributions with onsets at ±0.200 ± 0.001 V. For both the pixel-by-pixel numerical derivation (Figures

S18

S12b, h) and MP approaches (Figures 3b, c of main text and Figure S12i, n), the d$I$/d$V$ maps at -0.2 V and 0.2 V have a similar spatial structure, with a significant contribution from the center of the molecule. This is consistent with a molecular vibrational mode with an eigenenergy of 0.2 V (labelled VM1 in Figures S12b, h and i, n).

Fitting of the d$I$/d$V$ spectrum taken at the MgPc center (Figure 3a in main text; Figure S12a) also includes Fermi-Dirac distributions with onsets at ±0.089 ± 0.001 V (see above). The corresponding d$I$/d$V$ map at +0.09 V shows a dominant contribution from the center of the molecule (Figures S12g). Although our d$I$/d$V$ map at –0.09 V (Figure S12d) is dominated by strong d$I$/d$V$ signal at the peripheral Pc ligand, related to the degenerate, partially populated LUMOs (Gaussian peak at -0.097 ± 0.001 V in fits of d$I$/d$V$ spectra taken at MgPc periphery; Figure 3f of main text and solid filled blue curve in Figure S12a), it still shows a significant contribution from the center of the molecule, similar to the d$I$/d$V$ map at +0.09 V. We attribute these step-like features to vibrational modes associated with the metal-N bonds (VM2). This is consistent with similar d$I$/d$V$ maps observed by Mugarza *et. al.*[10b] for other metalated Pc's at comparable near-Fermi bias voltages.

Finally, the MgPc center d$I$/d$V$ spectra fits also include Fermi-Dirac distributions with onsets at ±0.005 ±0.001 V (Figure 3a in main text, Figure S12a; see above), that we attribute to other molecular vibrational modes (VM3) observed in similar experiments[12], as well as in surface enhanced Raman scattering measurements[9a] and in theoretical calculations[8a]. Note that the corresponding near-Fermi d$I$/d$V$ maps (Figures S12e, f, k, l) are again dominated by peripheral Pc contributions of the partially populated LUMOs at negative bias, and by the vibrational feature at positive bias (similar to that at ±0.09 V).

S19

## S9. d*I*/d*V* STS of single MgPc: partially populated LUMO and Hubbard *U* energy

Figure S13 shows d*I*/d*V* maps of a single MgPc, for bias voltages -0.05 to 0.5 V. The map at -0.05 V (Figure S13a) is dominated by contributions of the partially occupied, two-fold degenerate LUMOs, with its four-fold rotational symmetry and four nodal planes (grey arrows). For increasing bias voltage, d*I*/d*V* mapping evolves (Figures S13a to d) and, at 0.5 V, exhibits a spatial distribution similar to that of the LUMOs at -0.05 V. We ascribe this similarity and the respective bias voltage difference to a Hubbard *U* energy of ~0.55 eV required by an electron to overcome the Coulomb repulsion resulting from tunnelling into the partially occupied LUMOs. This is consistent with previous work on comparable systems[1a].

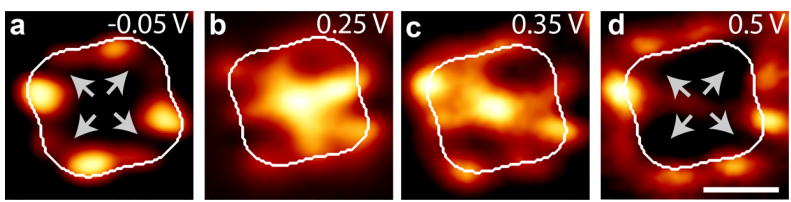

**Figure S13.** (a) – (d) d*I*/d*V* maps of single MgPc at $V_b$ = -0.05, 0.25, 0.35 and 0.5 V, respectively ($I_t$ = 250 pA; constant-current lock-in amplifier acquisition; see Methods in main text). Similarity between maps at -0.05 V (a) and 0.5 V (d) are indicative of partial population of the LUMOs, with an associated Hubbard *U* energy of ~0.55 eV. White curves: STM topographic molecular contours at corresponding bias. Scale bar: 1 nm.

# S10. d*I*/d*V* maps of "++" dimers with different intermolecular distances

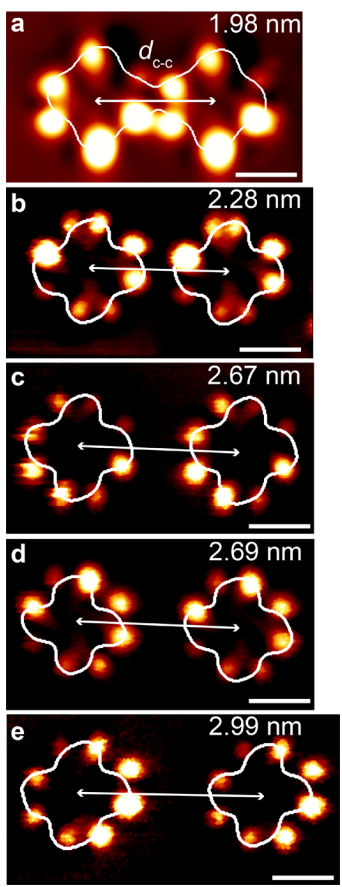

**Figure S14. MP d*I*/d*V* maps of MgPc "++" dimers with different Mg-Mg intermolecular separations $d_{c-c}$ ($V_b$ = -0.08 V).** White curves indicate MP STM topography molecular contours ($V_b$ = -2.5 V; $I_t$ = 100 pA; tip brought 150 pm closer to the surface for d*I*/d*V* acquisition). Scale bars: 1 nm.

# S11. Calculation of spectroscopic asymmetry parameters

The MgPc "++" dimer spectroscopic asymmetry parameters $\alpha_{dI/dV}$, $\alpha_{DFT}^{(Ag)}$, $\alpha_{DFT}^{(GP)}$, $\alpha_{LCMO}^{(Ag)}$, $\alpha_{LCMO}^{(GP)}$ in Figure 4a of the main text were calculated using the formula:

$$\alpha = \frac{\iint |\mathcal{R}_{90°}[(dI/dV)_{bin}(x,y,V_b = -0.08 \text{ V})] - (dI/dV)_{bin}(x,y,V_b = -0.08 \text{ V})| \, dx \, dy}{\iint |\mathcal{R}_{90°}[(dI/dV)_{bin}(x,y,V_b = -0.08 \text{ V})] + (dI/dV)_{bin}(x,y,V_b = -0.08 \text{ V})| \, dx \, dy}$$

where $(dI/dV)_{bin}(x,y,V_b = -0.08 \text{ V})$ is a d*I*/d*V* map at $V_b$ = -0.08 V after correction (Figures S15d, e, f) and binarization (Figure S15g, h, i), as a function of tip position (*x*, *y*), and $\mathcal{R}_{90°}[(dI/dV)_{bin}(x,y,V_b = -0.08 \text{ V})]$ is the 90° clockwise rotation of such map around an axis perpendicular to the surface going through the Mg center. The different parameters $\alpha_{dI/dV}, \alpha_{DFT}^{(Ag)}, \alpha_{DFT}^{(GP)}, \alpha_{LCMO}^{(Ag)}, \alpha_{LCMO}^{(GP)}$ were calculated using, respectively, experimental d$I$/d$V$ maps, DFT-calculated d$I$/d$V$ maps for MgPc on Ag(100), DFT-calculated d$I$/d$V$ maps for MgPc in the gas phase, d$I$/d$V$ maps for MgPc on Ag(100) calculated with our Linear Combination of Molecular Orbitals (LCMO) model (see below), and d$I$/d$V$ maps for MgPc in the gas phase calculated with our LCMO. The parameter $\alpha$ varies between 0 (when the Mg-Mg intermolecular distance $d_{c-c}$ within the "++" dimer is large and MgPc regains its four-fold rotational symmetry, as for the single MgPc case) and 1 (for the extreme case where the two MgPc's molecules within the "++" dimer are close to each other; see Figures S15e-l).

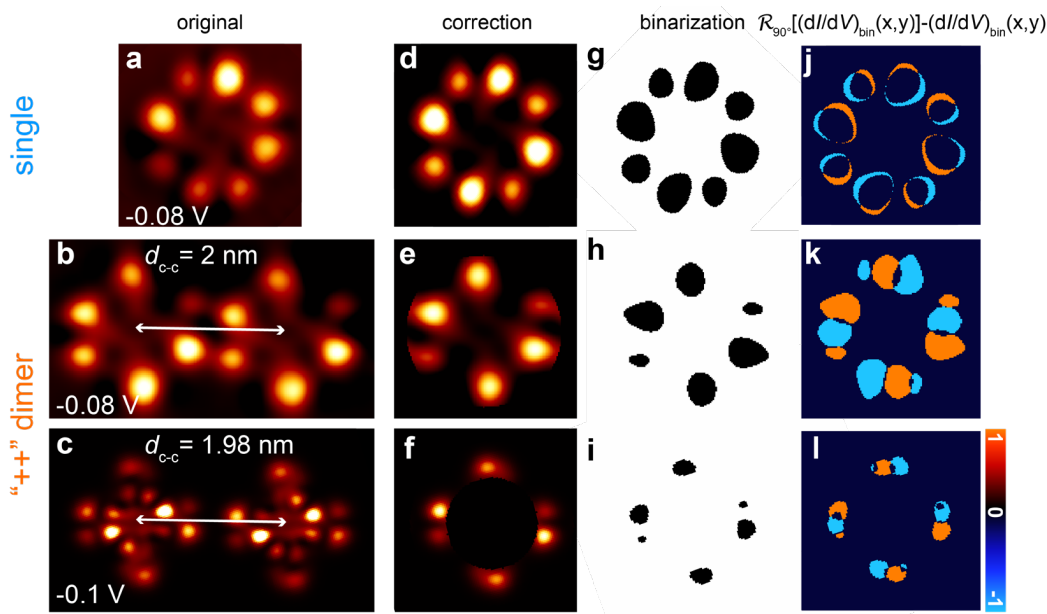

**Figure S15. Calculation of the spectroscopic asymmetry parameter $\alpha$.** (a) d$I$/d$V$ maps for single MgPc (experimental, $V_b$ = -0.08 V) and (b), (c) MgPc "++" dimer (experimental and DFT-calculated [see Methods in main text], respectively, $d_{c-c}$ = 1.98 nm, $V_b$ = -0.08 V). (d) – (f) Same maps as (a) – (c), for one MgPc, averaged with their 180° rotation around an axis perpendicular to the molecular plane and going through the Mg center, to correct for tip irregularities. For the DFT-simulated map, the intensity at the center of the molecule was removed to emphasize the break of rotational symmetry at the periphery. (g) – (i) Same maps as (d) – (f), binarized with a threshold defined at 75% of the maximum intensity. (j) – (l) Difference maps $\mathcal{R}_{90°}[(dI/dV)_{bin}(x,y)] - (dI/dV)_{bin}(x,y)$, where $(dI/dV)_{bin}(x,y)$ are the binarized maps in (g) – (i), and $\mathcal{R}_{90°}[(dI/dV)_{bin}(x,y)]$ are these maps rotated by 90° with respect to the axis going though Mg perpendicular to the molecular plane. These difference maps are used to calculate $\alpha$ according to the integral in the formula above.

## S12. Potential mechanisms behind the intermolecular interaction

Our experiments and DFT calculations show that intermolecular interactions are the cause of the observed reduction of molecular symmetry for MgPc in a "++" dimer. This break of symmetry is electronic in nature and involves the LUMOs. As discussed in section S1 above, CO-tip ncAFM measurements and DFT calculations indicate that the intramolecular morphology is identical for both single MgPc and "++" dimers (Figure S2); the observed reduction of symmetry is not the result of and is not accompanied by structural distortions. Below we discuss potential mechanims that could be behind the intermolecular interaction.

### a) Dipole-dipole interaction

As explained in the main text, MgPc adsorbed on Ag(100) is negatively charged due to surface-to-molecule electron transfer, with its partially filled LUMOs energy below the Fermi level. This results in a dipole moment formed by the negatively charged molecule and its positive image charge resulting from screening by Ag. In a MgPc "++" dimer, this results in two parallel dipole moments (Figure S16a), which could arguably give rise to a repulsive intermolecular interaction. In the following, we use first-order degenerate perturbation theory to estimate the possible effect of such repulsive dipole-dipole interaction on the LUMOs eigenenergies and spatial symmetries.

Let us consider a MgPc "++" dimer composed of MgPc molecules $A$ and $B$ with position vectors $\vec{r}_A$ and $\vec{r}_B$, i.e., intermolecular separation $d_{c-c} = \|\vec{r}_B - \vec{r}_A\|$ (Figure S16a). Based on the Bader analysis of our DFT calculations (see main text) and our experimental observation of partially populated LUMOs, we assume an Ag(100)-to-MgPc transfer of ~1 electron (charge $e$) per molecule, resulting in positive image charges and in electric dipole moments $|\vec{p}_A| = |\vec{p}_B| = |e| \cdot 2h$ for molecules $A$ and $B$, where $h = 2.6$ Å (extracted from DFT calculations of the relaxed molecular structure, see Figure S2) is the molecular adsorption height. This can give rise to a repulsive intermolecular dipole-dipole interaction $H_{\text{dipole}}(\vec{r}_A, \vec{r}_B) = H_{\text{dipole}}(d_{\text{c-c}}) = \frac{|\vec{p}_A| \cdot |\vec{p}_B|}{4\pi\varepsilon_0} \frac{1}{\|\vec{r}_B - \vec{r}_A\|^3} = \frac{|\vec{p}_A| \cdot |\vec{p}_B|}{4\pi\varepsilon_0} \frac{1}{d_{\text{c-c}}^3}$.

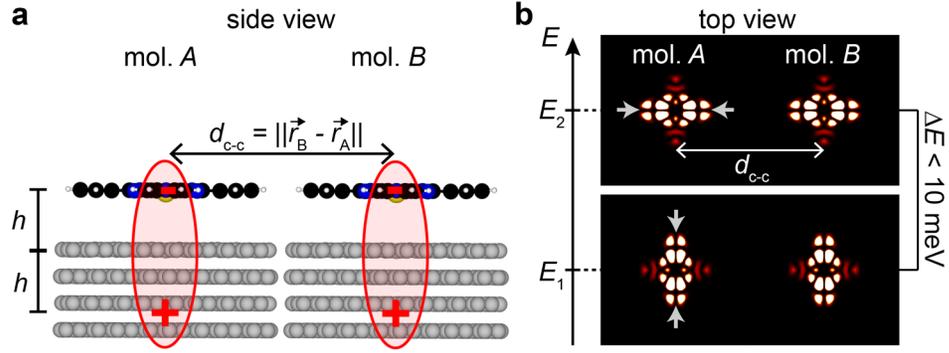

**Figure S16 Repulsive dipole-dipole interaction between MgPc molecules in a "++" dimer.** (a) Side view of the DFT-relaxed MgPc molecules $A$ and $B$ on Ag(100). Upon adsorption, each molecule becomes negatively charged due to a transfer of an electron from the surface, resulting in a positive image charge, and hence a dipole. In an MgPc "++" dimer with Mg-Mg distance $d_{\text{c-c}}$, this can result in a repulsive dipole-dipole interaction. (b) Top view of the lowest (below) and highest (above) eigenenergy LUMOs (modulus squared; horizontal cut 1.7 Å above the molecular plane), where the energy degeneracy is lifted by the repulsive dipole-dipole interaction. We calculated these LUMOs via first-order degenerate perturbation theory, by considering the two-fold degenerate unperturbed LUMOs of isolated gas phase MgPc given by DFT.

Let us consider the effect of this hypothetical dipole-dipole interaction on the LUMOs of molecule $A$, which, when $d_{\text{c-c}} \to \infty$, are two-fold degenerate. According to first-order degenerate perturbation theory[13], and in the basis set $\left\{\left|LUMO_1^{(0)}\right\rangle; \left|LUMO_2^{(0)}\right\rangle\right\}$ defined by these two unperturbed LUMOs, we can write the perturbed LUMOs of $A$ as $|LUMO(d_{\text{c-c}})\rangle = c_1(d_{\text{c-c}})\left|LUMO_1^{(0)}\right\rangle + c_2(d_{\text{c-c}})\left|LUMO_2^{(0)}\right\rangle$.

The coefficients $c_l(d_{\text{c-c}})$ ($l$ = 1, 2) satisfy the Schrödinger equation:

$$\begin{pmatrix} H_{11}(d_{\text{c-c}}) & H_{12}(d_{\text{c-c}}) \\ H_{21}(d_{\text{c-c}}) & H_{22}(d_{\text{c-c}}) \end{pmatrix} \cdot \begin{pmatrix} c_1(d_{\text{c-c}}) \\ c_2(d_{\text{c-c}}) \end{pmatrix} = E(d_{\text{c-c}}) \begin{pmatrix} c_1(d_{\text{c-c}}) \\ c_2(d_{\text{c-c}}) \end{pmatrix}$$

with

$$H_{lm}(d_{c-c}) = \left\langle LUMO_l^{(0)} \middle| \left(H_0 + H_{\text{dipole}}(d_{c-c})\right) \middle| LUMO_m^{(0)} \right\rangle$$

$$H_0 \left| LUMO_l^{(0)} \right\rangle = E_0 \left| LUMO_l^{(0)} \right\rangle$$

where $l, m = 1, 2$. $E_0$ is the eigenenergy of the unperturbed two-fold degenerate LUMOs (from our experiment, $E_0 = -0.097 \pm 0.001$ eV with respect to the Fermi level) and $E(d_{c-c})$ is the eigenenergy of the LUMOs perturbed by the dipole-dipole interactions.

We find $E(d_{c-c})$ by solving:

$$\begin{vmatrix} H_{11}(d_{c-c}) - E(d_{c-c}) & H_{12}(d_{c-c}) \\ H_{21}(d_{c-c}) & H_{22}(d_{c-c}) - E(d_{c-c}) \end{vmatrix} = 0$$

yielding two values $E_l(d_{c-c})$ ($l$ = 1, 2) of $E(d_{c-c})$ for a given $d_{c-c}$, and for each value of $E_l(d_{c-c})$ an associated perturbed $|LUMO_l(d_{c-c})\rangle$. That is, the dipole-dipole interaction can lift the LUMOs degeneracy: $E_1(d_{c-c}) \neq E_2(d_{c-c})$.

By considering the two-fold degenerate LUMOs of the isolated gas-phase MgPc given by DFT (see Figure S17a), we obtain $\Delta E(d_{c-c}) = |E_2(d_{c-c}) - E_1(d_{c-c})| < 10$ meV, even for limits of unphysically close distances $d_{c-c}$ and unphysically large values of the dipole moment. We further estimate that any possible interaction due to induced in-plane dipoles (London dispersion[14] with energy potential scaling as $\propto d_{c-c}^{-6}$) must be even smaller. Note that in our experiments, we observed energy differences between the lowest and 2nd lowest eigenenergy LUMOs of ~0.25 eV (Figures 3f, i, j of main text).

Figure S16b shows the modulus squared of the resulting perturbed, non-degenerate wavefunctions of molecules $A$ and $B$ (cut 1.7 Å above the molecular plane), for $d_{c-c} \approx 1$ nm, determined by considering the gas phase unperturbed $\left|LUMO_1^{(0)}\right\rangle$ and $\left|LUMO_2^{(0)}\right\rangle$ of isolated MgPc calculated by DFT (Figure S17a). It is important to note that the LUMO with the lowest (highest) eigenenergy has a nodal plane (gray arrows in Figure S16b) perpendicular (parallel, respectively) to the intermolecular Mg-Mg axis of the MgPc "++" dimer. In our experiments (see Figures 3i, j of main text) and DFT calculations (Figures 3m, n of main text), we observed exactly the opposite, i.e., lowest (highest) eigenenergy orbital with nodal plane parallel (perpendicular) to the Mg-Mg axis. This means that the lift of LUMOs degeneracy and reduction of LUMOs rotational symmetry observed in our experiments and DFT calculations are not the result of a repulsive dipole-dipole interaction. Notably, this demonstrates that such lift of LUMOs degeneracy and reduction of LUMOs symmetry must be the result of an effectively attractive MgPc-MgPc interaction.

**b)        Interaction mediated by Friedel oscillations of Ag(100) bulk electron density**

In the following, we discuss the possibility that the effectively attractive MgPc-MgPc interaction in a "++" dimer is mediated by Friedel oscillations of the Ag(100) bulk electron density, that could result from the presence of the molecular adsorbates, similar to the case of metal adatoms[15]. The energy of such a substrate-mediated interaction would be oscillatory, with a period equal to half the Fermi de Broglie wavelength[16] of bulk Ag. That is, the interaction would alternate between repulsive, null (in the equilibrium case) or attractive, depending on $d_{c-c}$, with a period[17] in the (100) plane of Ag of, at most, ~2.5 Å. Note that Ag(100) does not host Shockley surface state electrons; these states can give rise to electron density Friedel oscillations with a significantly larger periodicity[18]. In our experiments and DFT calculations, we observed that the lowest (highest) eigenenergy LUMO in a "++" dimer always has a nodal plane parallel (perpendicular, respectively) to the Mg-Mg axis (Figures 3i, j, m, n of main text), with $\alpha_{dI/dV}$ and $\alpha_{DFT}^{(Ag)}$ decaying monotonically (without oscillations) as a function of $d_{c-c}$ (Figure 4a in main text). This is an indication that the effective intermolecular interaction is attractive (or null when $d_{c-c}$ becomes large); if it were repulsive for some specific values of $d_{c-c}$, the break of LUMO symmetry would be opposite, with the lowest (highest) eigenenergy LUMO having a nodal plane perpendicular (parallel, respectively) to the Mg-Mg axis (see effect of repulsive dipole-dipole interaction above). We therefore conclude that the observed lift of LUMOs degeneracy and reduction of LUMOs symmetry cannot be the result of an oscillatory Friedel interaction mediated by the bulk Ag(100) electrons.

c) **Ruderman-Kittel-Kasuya-Yosida (RKKY) interaction between unpaired spins**

We concluded that the LUMOs of MgPc are partially filled due to Ag(100)-to-molecule electron transfer (see main text). In a MgPc "++" dimer, electrons that partially populate these LUMOs and remain unpaired can potentially give rise to Ruderman-Kittel-Kasuya-Yosida (RKKY) spin-spin interactions mediated by Ag(100) bulk electrons[19]. Similar to the interaction that can be mediated by Friedel oscillations of the Ag(100) bulk electron density (see above), the RKKY interaction potential $U_{\text{RKKY}}(r)$ oscillates with interspin distance $r$ (i.e., can alternate between attractive or repulsive)[19a]:

$$U_{\text{RKKY}}(r) = \frac{16 J_{\text{eff}}^2 2 m_e k_F^4}{(2\pi)^3 \hbar^2} \left( \frac{\cos(2k_F r)}{(2k_F r)^3} - \frac{\sin(2k_F r)}{(2k_F r)^4} \right)$$

where $k_F$ is the Fermi wavenumber of the Ag(100) bulk electrons, $m_e$ is the electron mass and $J_{\text{eff}}$ describes the coupling between the spin at the adsorbate and the substrate bulk electrons. Here, as an example, we assume for $J_{\text{eff}}$ the hypothetical case of a $d$-electron spin (however, note that, given the reduced localisation of the additional charge in the molecular ligand, we expect an even smaller effect)[19b]:

$$J_{\text{eff}} = -\frac{\Delta}{\pi \cdot \text{DOS}(E_0)} \cdot \frac{U}{|E_0| \cdot (U - |E_0|)}$$

where $\text{DOS}(E_0)$ is the bulk Ag(100) density of electronic states at eigenenergy $E_0 \cong -0.1$ V (with respect to the Fermi level) of the considered unpaired adsorbate electron (here, we estimated $\text{DOS}(E_0)$ at the Fermi level), $\Delta \cong 0.15$ V is the FWHM of the occupied LUMO (estimated from our experimental d$I$/d$V$ data), and $U \cong 0.55$ V the Hubbard energy[20] (lower limit estimated from our STS d$I$/d$V$ mapping; see Figure S13). Using these expressions, we estimate $U_{\text{RKKY}} \approx 9$ meV for $r \approx 5$ Å (which is close to

S27

the distance between adjacent isoindole groups in a DFT-relaxed MgPc "++" dimer with the smallest $d_{c-c}$), significantly smaller that the characteristic energies involved in the observed lift of LUMOs degeneracy. Combined with its oscillatory nature (see reasons invoked above for Friedel oscillations), we exclude the possibility that an RKKY interaction is the cause of the lift of LUMOs degeneracy and reduction of LUMOs symmetry.

## S13. Electronic structure of single MgPc: DFT calculations

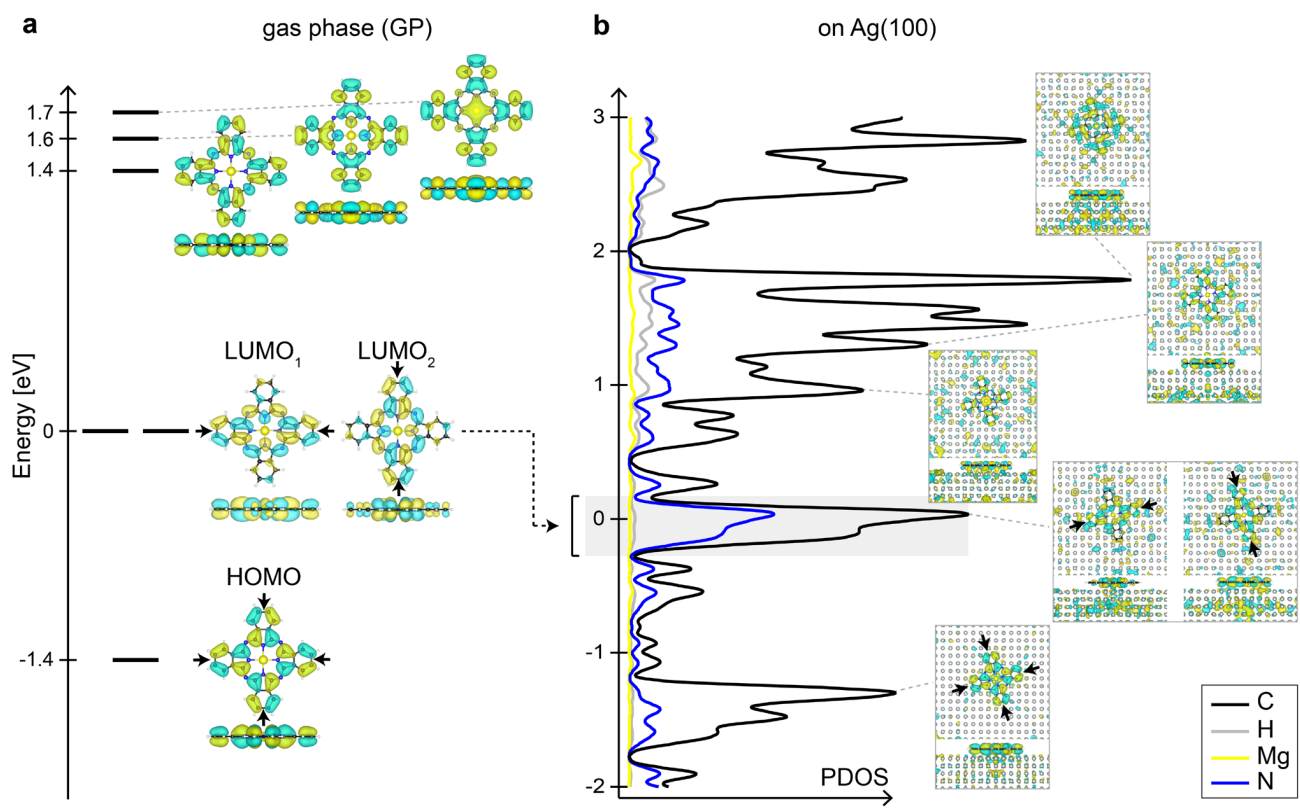

**Figure S17.** DFT-calculated electronic energy level diagram [in (b), projected density of states (PDOS)] and orbital wavefunction isosurfaces (top and side views) for a single MgPc (a) in the gas phase (i.e., neutrally charged) and (b) on Ag(100). Wavefunction isosurface values: (a) 0.013 $a_0^{-3/2}$; (b) 0.07 $a_0^{-3/2}$ ($a_0$: Bohr radius). Different colors correspond to different signs of the wavefunction. Black arrows indicate nodal planes of HOMO and LUMOs. Energy referenced to LUMOs (a) and to Fermi level (b). Note spectral energy broadening and spatial delocalization of molecular orbitals upon adsorption on Ag(100).

## S14. Linear combination of molecular orbitals (LCMO) model

We used a simple model to provide insight into the observed lift of LUMOs degeneracy and break of LUMOs rotational symmetry in the MgPc "++" dimer. This model is based on the hypothesis of hybridization between delocalized LUMO/Ag(100) orbitals from the two different MgPc's of a "++" dimer, where new hybrid orbitals result from a linear combination of such delocalized LUMO/Ag(100) states (LCMO). We assume that the MgPc negative charge resulting from partial filling of the LUMOs due to surface-to-molecule electron transfer is screened by the Ag(100) conduction electrons, effectively cancelling intermolecular Coulomb repulsion.

Let us consider a MgPc "++" dimer on Ag(100) composed of MgPc molecules $A$ and $B$ with Mg center position vectors $\vec{r}_A$ and $\vec{r}_B$, i.e., intermolecular separation $d_{c-c} = \|\vec{r}_B - \vec{r}_A\|$. Let us first assume that the two MgPc molecules are not interacting with each other: LUMOs $|LUMO_{A1}\rangle$ and $|LUMO_{A2}\rangle$ of molecule $A$, and $|LUMO_{B1}\rangle$ and $|LUMO_{B2}\rangle$ of molecule $B$ are degenerate, with eigenenergy $E_0 \cong -0.1 \text{ eV} - \phi_{Ag(100)}$ (i.e., referenced with respect to the vacuum level here, where $\phi_{Ag(100)} = 4.26 \text{ eV}$ is the Ag(100) work function[21]). In the following, we will use as $|LUMO_{A1}\rangle$, $|LUMO_{A2}\rangle$, $|LUMO_{B1}\rangle$ and $|LUMO_{B2}\rangle$ the DFT-calculated states associated with the two-fold degenerate LUMOs of a single, DFT-relaxed MgPc adsorbed on Ag(100) (i.e., unperturbed by the presence of another MgPc molecule in proximity). Due to hybridisation with Ag(100) conduction electrons, these orbitals are significantly delocalized (see Figure S17b). Figure S18a shows $|LUMO_{A1}(x,y,z)|^2 + |LUMO_{A2}(x,y,z)|^2 = |\langle x,y,z|LUMO_{A1}\rangle|^2 + |\langle x,y,z|LUMO_{A2}\rangle|^2$, for a single, DFT-relaxed MgPc on Ag(100), for $z = 1.7$ Å (with respect to the molecular plane).

Now, let us consider the MgPc "++" dimer system, with both molecules $A$ and $B$ interacting with each other, with Hamiltonian $H$. We are interested in solutions $|\varphi\rangle$ of the Schrödinger equation $H|\varphi\rangle = E|\varphi\rangle$ where $|\varphi\rangle$ is written as a linear combination of unperturbed orbitals $|LUMO_{A1}\rangle$, $|LUMO_{A2}\rangle$, $|LUMO_{B1}\rangle$ and $|LUMO_{B2}\rangle$ : $|\varphi\rangle = c_{A1}|LUMO_{A1}\rangle + c_{A2}|LUMO_{A2}\rangle + c_{B1}|LUMO_{B1}\rangle + c_{B2}|LUMO_{B2}\rangle$. This is a supramolecular analogue of a linear combination of atomic orbitals model for molecular orbitals. The Schrödinger equation becomes:

$$\begin{pmatrix} H_{A1,A1} & H_{A1,A2} & H_{A1,B1} & H_{A1,B2} \\ H_{A2,A1} & H_{A2,A2} & H_{A2,B1} & H_{A2,B2} \\ H_{B1,A1} & H_{B1,A2} & H_{B1,B1} & H_{B1,B2} \\ H_{B2,A1} & H_{B2,A1} & H_{B2,B1} & H_{B2,B2} \end{pmatrix} \begin{pmatrix} c_{A1} \\ c_{A2} \\ c_{B1} \\ c_{B2} \end{pmatrix} = E \begin{pmatrix} c_{A1} \\ c_{A2} \\ c_{B1} \\ c_{B2} \end{pmatrix}$$

with matrix elements $H_{l,l} = E_0$ and $H_{l,m} = \langle LUMO_l|H|LUMO_m\rangle$ when $l \neq m$ ($l, m = A1, A2, B1, B2$). Eigenenergies $E$ are determined by solving:

$$\begin{vmatrix} E_0 - E & H_{A1,A2} & H_{A1,B1} & H_{A1,B2} \\ H_{A2,A1} & E_0 - E & H_{A2,B1} & H_{A2,B2} \\ H_{B1,A1} & H_{B1,A2} & E_0 - E & H_{B1,B2} \\ H_{B2,A1} & H_{B2,A1} & H_{B2,B1} & E_0 - E \end{vmatrix} = 0$$

We assume a simple model Hamiltonian $H = -\frac{\hbar^2}{2m}\nabla^2 - \frac{\beta e^2}{4\pi\epsilon_0}\left(\frac{1}{\|\vec{r}-\vec{r}_A\|} + \frac{1}{\|\vec{r}-\vec{r}_B\|}\right)$. That is, it is implied here that the additional negative charge that each MgPc of the "++" dimer gained due to Ag(100)-to-molecule electron transfer is screened by the metal substrate conduction electrons, and that, at the same time, the nuclei of each molecule contribute to an effective attractive $\propto -1/r$ potential centered at the Mg atom. We fixed constant $\beta \cong 5.9$ in $H$ such that, for the MgPc "++" dimer with the smallest Mg-Mg distance observed in our experiments (i.e., $d_{c-c} \cong 2$ nm), the difference $|E_2 - E_1|$ between the two smallest eigenenergies $E$ is ≈ 250 meV, i.e., the experimental energy difference between highest and lowest energy LUMO $dI/dV$ peak for such a dimer (see Figure 3a of main text).

Solving the matrix Schrödinger equation above yields a set of coefficients $c_l$ ($l = A1, A2, B1, B2$) for each eigenenergy $E$. We focus on the two non-degenerate wavefunctions $|\varphi_1\rangle$ and $|\varphi_2\rangle$ associated with the two smallest eigenenergies $E_1$ and $E_2$. Figure S18b shows $|\varphi_1(x, y, z)|^2 = |\langle x, y, z|\varphi_1\rangle|^2$ (associated with the lowest eigenenergy $E_1$) for an MgPc "++" dimer with $d_{c-c} = 2.12$ nm, for $z = 1.7$ Å (with

respect to the molecular plane). This low-energy $|\varphi_1|^2$ exhibits a break of four-fold rotational symmetry in comparison with the single MgPc $|LUMO_{A1}|^2 + |LUMO_{A2}|^2$ in Figure S18a, with a nodal plane parallel to the Mg-Mg axis (grey arrows).

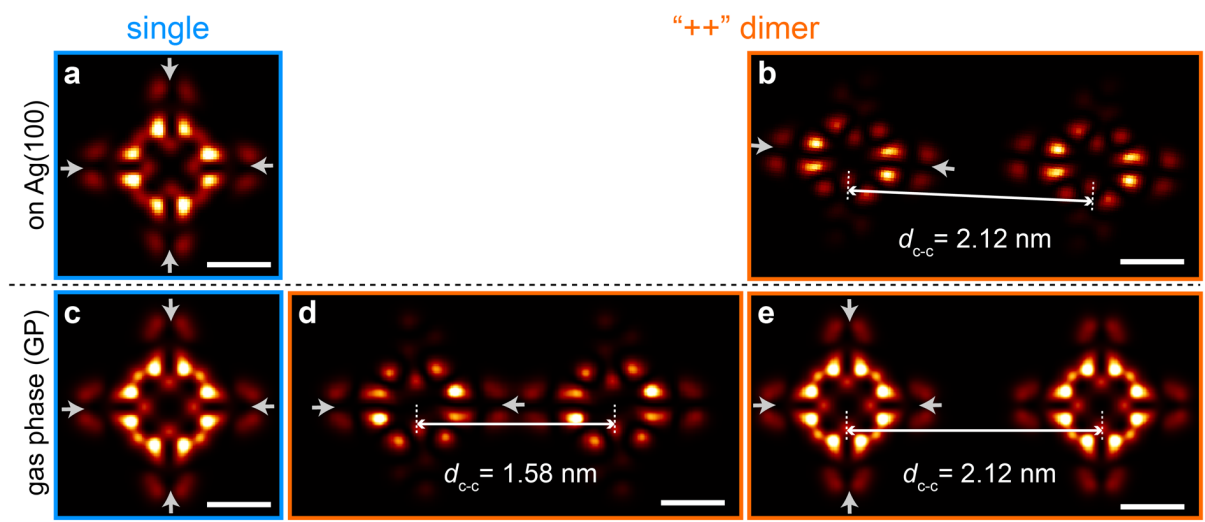

**Figure S18. Hybrid orbitals given by our linear combination of molecular orbitals (LCMO) model, on Ag(100) and in the gas-phase.** (a) Top view of $|LUMO_{A1}(x,y,z)|^2 + |LUMO_{A2}(x,y,z)|^2$ for single DFT-relaxed MgPc on Ag(100), with two-fold degenerate $|LUMO_{A1}\rangle$ and $|LUMO_{A2}\rangle$ calculated by DFT. (b) Top view of low-energy LUMO $|\varphi_1(x,y,z)|^2 = |\langle x,y,z|\varphi_1\rangle|^2$ resulting from our LCMO model, for MgPc "++" dimer on Ag(100), with $d_{c-c} = 2.12$ nm. (c) Same as (a), for single DFT-relaxed MgPc in the gas phase. (d), (e) Same as (b), for MgPc "++" dimers in the gas phase, with $d_{c-c} = 1.58$ and 2.12 nm, respectively. Grey arrows indicate wavefunction nodal planes. All maps sampled at $z = 1.7$ Å above the molecule plane. Scale bar: 1 nm (applies to all panels).

For comparison, we performed the same calculation as above, but for a single neutral MgPc and for neutral MgPc "++" dimers in the gas phase. Figure S18c shows the DFT-calculated $|LUMO_{A1}(x,y,z)|^2 + |LUMO_{A2}(x,y,z)|^2 = |\langle x,y,z|LUMO_{A1}\rangle|^2 + |\langle x,y,z|LUMO_{A2}\rangle|^2$, for a single, DFT-relaxed MgPc in the gas-phase. Figures S18d, e show results of our LCMO model applied to interacting gas-phase MgPc "++" dimers, with $d_{c-c} = 1.58$ and 2.12 nm, respectively. A break of MgPc orbital four-fold rotational symmetry is also observed, but only for a value of $d_{c-c}$ significantly smaller

S31

than for the substrate-hybridized orbitals in Figure S18b; the gas phase MgPc LUMOs become again degenerate and regain four-fold symmetry for $d_{c-c} \cong 2$ nm (Figure S18e).

Figure S19 shows theoretically calculated differential conductance maps (see Methods in main text) of MgPc "++" dimers with different values of $d_{c-c}$, on Ag(100) and in the gas phase, at an energy associated with the low-energy LUMO (Figures S19a, c: fully calculated with DFT; Figures S19b, d: generated with our LCMO model). Results of our DFT calculations and LCMO model are consistent with each other, with the low-energy LUMO exhibiting a break of four-fold rotational symmetry for $d_{c-c}$ up to ~3 nm on Ag(100) (in quantitative agreement with our experimental data), and for $d_{c-c} < 2$ nm in the gas phase. The LUMO delocalization resulting from hybridization with Ag(100) electronic states gives rise to an intermolecular interaction within the MgPc "++" dimer that is of significantly longer range than in the gas phase.

It is important to note that our simple LCMO model, by construction, produces four energetically distinct states $|\varphi_j\rangle$ (with $j = 1…4$). Experimentally, we only identified two of these states (Figures 3i, j of main text), that we associate with $|\varphi_1\rangle$ and $|\varphi_2\rangle$. We speculate that the "missing two" ($|\varphi_3\rangle$, $|\varphi_4\rangle$) reside in an energy range where the molecular d$I$/d$V$ spectral signature is more complex and is dominated by other effects (e.g., vibrational mode at ~0.2 V).

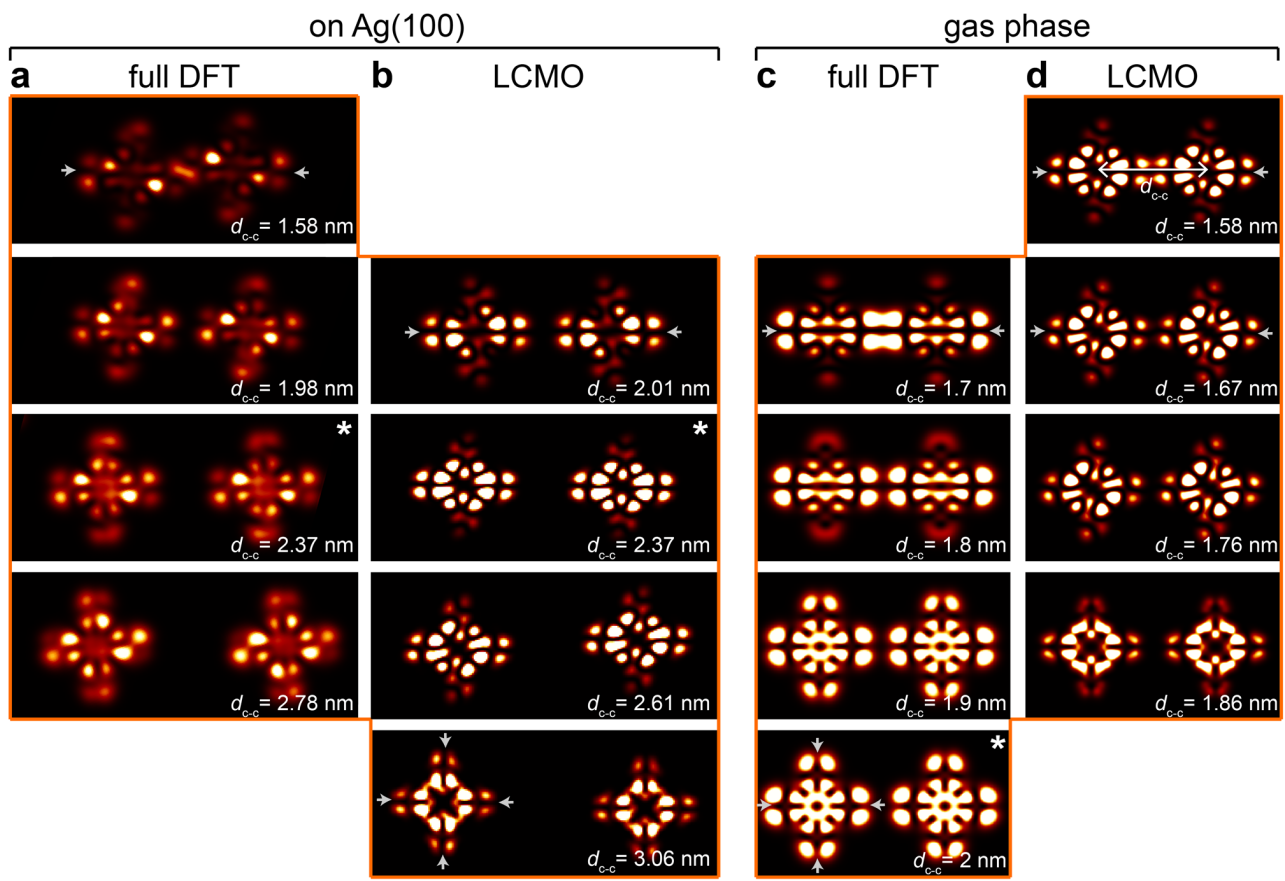

**Figure S19. Theoretical differential conductance maps associated with low-energy LUMO of MgPc "++" dimer on Ag(100) and in the gas phase: DFT calculations and LCMO model.** (a) Top view of simulated d$I$/d$V$ maps corresponding to the DFT-calculated low-energy LUMO of MgPc "++" dimers on Ag(100), for different values of $d_{c-c}$. (b) Top view of simulated d$I$/d$V$ maps corresponding to the low-energy LUMO $|\varphi_1(x,y,z)|^2 = |\langle x,y,z|\varphi_1\rangle|^2$ resulting from our LCMO model, for MgPc "++" dimers on Ag(100) with different values of $d_{c-c}$. (c), (d) Same as (a) and (b), respectively, but for MgPc "++" dimers in the gas phase. A reduction of LUMO rotational symmetry (from four-fold to two-fold) is observed on Ag(100) for $d_{c-c}$ up to ~3 nm. A reduction of LUMO rotational symmetry is also observed in the gas phase, but for significantly smaller values of $d_{c-c}$ (< ~ 2 nm; e.g., compare maps labelled with '*'). Grey arrows indicate wavefunction nodal planes. All maps sampled at $z = 4.3$ Å above the molecule plane (see Methods in main text).